\documentclass[twocolumn,twocolappendix]{aastex63}
\usepackage{physics}
\usepackage{graphicx}
\usepackage[version=4]{mhchem}
\usepackage{amsmath}
\usepackage{hyperref}

\newcommand{\cmsg}{cm$^{2}\cdot$ g$^{-1}$}
\newcommand{\gcmc}{g$\cdot$cm$^{-3}$}

\newcommand{\kms}{km$\cdot$s$^{-1}$}
\newcommand{\msunyear}{M$_{\odot}\cdot$year$^{-1}$}
\newcommand{\msun}{M$_{\odot}$}
\newcommand{\rsun}{R$_{\odot}$}
\newcommand{\lsun}{L$_{\odot}$}

\graphicspath{{./}{figures/}}

\newcommand{\new}[1]{{\color{black}}#1} 
\newcommand{\newr}[1]{{\color{black}}#1} 
\newcommand{\newrr}[1]{{\color{black}#1}} 

\begin{document}

\defcitealias{chen2017}{C17}

\title{A 3D radiation-hydrodynamic AGB binary model}

\correspondingauthor{Zhuo Chen}
\email{zc10@ualberta.ca}

\correspondingauthor{Natalia Ivanova}
\email{nata.ivanova@ualberta.ca}

\author[0000-0001-7420-9606]{Zhuo Chen}
\affiliation{Department of Physics, University of Alberta, Edmonton, AB T6G 2E1, Canada}
\affiliation{CITA National Fellow}

\author{Natalia Ivanova}
\affiliation{Department of Physics, University of Alberta, Edmonton, AB T6G 2E1, Canada}

\author{Jonathan Carroll-Nellenback}
\affiliation{Department of Physics and Astronomy, University of Rochester, Rochester, NY 14627, USA}



\begin{abstract}
The origin of chemically peculiar stars and non-zero eccentricity in evolved close binaries have been long-standing problems in binary stellar evolution. Answers to these questions may trace back to an intense mass transfer during AGB binary phase.
In this work, we use {\tt AstroBEAR} to solve the 3D radiation-hydrodynamic equations and calculate the mass transfer rate in asymptotic-giant-branch (AGB) binaries that undergo the wind-Roche-lobe-overflow or Bondi-Hoyle-Lyttleton (BHL) accretion. {\tt MESA} produces the density and temperature of the boundary condition of the AGB star. To improve the resolution of the dynamics of a circumbinary disk, we implement an azimuthal angle-dependent 3D radiation transfer. We consider optically thin cooling and obtain the number density of the coolants by solving Saha equations. One of the goals of this work is to illustrate the transition from the wind-Roche-lobe-overflow to BHL accretion. 
Both circumbinary disks and spiral structure outflows can appear in the simulations.
Circumbinary disks may form when the optical thickness in the equatorial region increases. The increase of the optical thickness is due to the deflected wind.
The resulting mass transfer efficiency in our models is up to a factor of 8 times higher than what the standard BHL accretion scenario predicts, and the outflow gains up to $91\%$ of its initial angular momentum when it reaches 1.3 binary separations.
Consequently, some AGB binaries may undergo orbit shrinkage, and some will expand.
The high mass transfer efficiency is closely related to the presence of the circumbinary disks.
\end{abstract}


\section{Introduction}

Asymptotic-giant-branch (AGB) stars have a significantly larger size ($\sim1$AU) than their main-sequence (MS) counterparts. They have pulsating atmosphere \citep{vlemmings2017,khouri2019} and may exhibit variability with long periods ranging from 200 days to 1000 days \citep{mowlavi2018,karambelkar2019}. AGB stars are one of the major sites in galaxies that produce metal. Metal can be carried away from the AGB stars by radiation-driven AGB winds when dust forms. \new{The speed of the AGB wind varies from $4-20$\ \kms \citep{hofner2018}, and a companion star may capture the wind with its gravity.} In the case that there is a MS star close to an AGB star, a substantial fraction of the mass-loss may be accreted onto the MS companion \citep{chen2017,saladino2018,saladino2019a}. As a result, the metallicity of the companion may change. Such early-stage low-mass stars become chemically peculiar, and their future evolution will be strongly affected. Carbon-enhanced-metal-poor (CEMP) stars \citep{beers2005,abate2013,abate2015}, Barium stars \citep{bidelman1951,escora2019}, CH stars \citep{keenan1942,mcclure1990} and dwarf carbon stars \citep{dahn1977,roulston2019} are common examples of the chemically peculiar stars. Their existence could be the evidence of the mass transfer during the previous AGB binary phase. The binarity of CH stars and CEMP stars has been studied \citep{mcclure1990,starkenburg2014,jorissen2016}, confirming that many of them have companions. \new{A number of recent studies show that the eccentricity of some of the aforementioned chemically peculiar stars may be large \citep{hansen2016,jorissen2016,vanderswaelmen2017,oomen2018,jorissen2019}, and their orbital periods ranging from hundreds to thousands of days.} The non-zero eccentricity in these close binary stars indicates some intense interactions that can pump the eccentricity may happen during their AGB binary phases. A strong correlation between a circumstellar disk and binarity has also been established in galactic RV Tauri stars \citep{manick2017}. Furthermore, many RV Tauri stars show a lack of refractory elements, which is called 'depletion' \citep{giridhar1994,vanwinckel1998}. Some researches suggest that the reaccretion of gas from a circumstellar disk around the post-AGB star \citep{gezer2019,oomen2019} may induce the 'depletion'. Besides the 'smoking gun' evidences, observations also reveal that dusty circumbinary disks exist in binary systems with evolved stars \citep{kervella2015,hillen2016,homan2017,ertel2019}. UV excess of some AGB stars also suggests that there could be accreting MS companions near them \citep{sahai2008,ortiz2016}.

\begin{figure}
    \centering
    \includegraphics[width=\columnwidth]{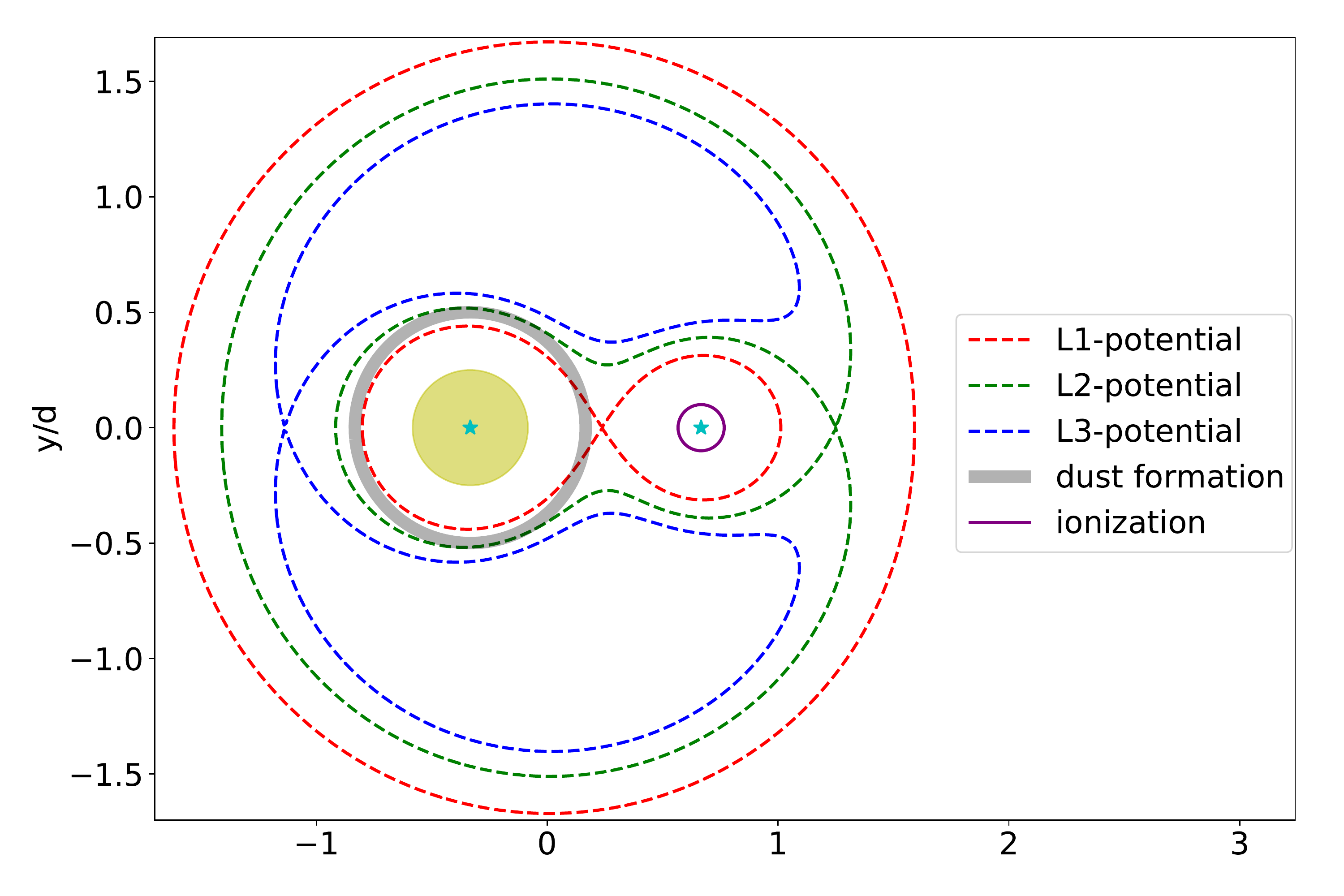}
    \caption{A schematic picture of the WRLOF. The coordinate is scaled to the binary separation $d$. The contours of the L1, L2, and L3 Lagrangian potential are drawn in red, green, and blue dashed lines. The AGB star is in yellow color, which is filling a fraction of its Roche lobe. Its dust forming region is represented by the gray ring. The companion star may have an ionization region, and we indicate that region by a purple line.}
    \label{fig:wrlof}
\end{figure}

In close AGB binaries, the mass transfer efficiency almost determines the intensity of the interaction between the two stars. However, it has been a difficult problem to quantify the mass transfer efficiency observationally because of the uncertainty in the measurement of the mass-loss rate and accretion rate. \new{Fortunately, with the rapidly increasing computational power, it is now possible to quantify the mass transfer efficiency in AGB binaries by carrying out 3D global numerical calculations.} Figure \ref{fig:wrlof} shows a schematic and global picture of an AGB binary undergoing wind-Roche-lobe-overflow (WRLOF) as described in \citet{Podsiadlowski2007}.
The main difference between the WRLOF and Roche-lobe overflow (RLOF) is that the AGB star is not filling its Roche-lobe to the full. \new{Therefore, the binary separation of an AGB binary undergoing WRLOF should be larger than the binary separation if it is undergoing RLOF.} On the other hand, if the binary separation is so significant that not only the physical size of the AGB star but also its dust formation region is much smaller than the Roche lobe of the AGB star, the mass transfer mechanism approaches the Bondi-Hoyle-Lyttleton (BHL) accretion limit \citep{hoyle1939,bondi1944,edgar2004}. In that scenario, the radiation driven AGB wind becomes supersonic before it crosses the Roche lobe of the AGB star. The two stars become not causally connected by hydrodynamics. The secondary may accrete mass from the AGB wind without imposing much effect on the AGB star's atmosphere. Such a scenario has been carefully investigated in simulations that focus on the hydrodynamics around the secondary \citep{huarte2013}. Although we qualitatively know that WRLOF is different from the BHL accretion \citep{mohamed2012}, the transition from the WRLOF to the BHL accretion is still poorly understood. One goal of this research is to illustrate such a transition with incremental change of the binary separation.

Dust formation is the key to model the mass transfer process in AGB binaries. When dust forms, the opacity of the fluid may increase by a factor of $10^{4}$, and the outflow may become supersonic in a short time under the radiation pressure. The location of the dust formation region almost determines the domain of dependence (circumbinary disk scenario is an exception) of both stars and the mass transfer mechanism. The dust formation problem has been studied from a kinetic perspective by modeling the dust growth and sublimation from clusters of small particles \citep{gail1984,gail1985,gail1986,gail1988,gail1999}. The method has been used by \citet{helling2006} to study the dust in brown dwarfs and applied to the AGB wind problem in 1D \citep{jeong2003}, 2D \citep{woitke2006a,woitke2006b}, and 3D \citep{freytag2008}. \new{More recently, \citet{hofner2019} simulated M-type AGB stars in 3D with their new kinetic dust formation model \citep{hofner2016}.} The first principle approaches are invaluable because the dust formation and destruction are susceptible to the radiation and hydrodynamical processes. However, an accurate (and sensitive) dust formation model would intrinsically require other parts of physics such as chemistry, radiation transfer, and shock physics to be as accurate.

The chemistry plays a vital role in the formation of dust and the dynamics of AGB winds. \citet{boulangier2019a} modeled the AGB atmospheres and the AGB winds in a dust-free and radiation-free environment. They emphasize the chemical non-equilibrium nature around the AGB stars and conclude that the chemical processes may regulate the thermodynamics of the AGB stars' atmosphere, thus strongly affect the properties of the winds. Recently, \citet{boulangier2019b} endeavored to model the onset of nucleation process from the non-equilibrium chemical reaction. The nucleation process may explain the origin of the small particles in the kinetic model. However, the biggest obstacle of such an approach is the lack of quantitative information about some reaction rates.

The radiation in the atmosphere of an AGB star may not be in local thermal equilibrium (LTE) with the gas, and the scattering of the radiation may be non-negligible. \citet{woitke2006a} solved the radiative transfer problem in their 2D model with the Monte Carlo method. Although \citet{woitke2006a} reported a negative result on the formation of the dusty wind, they pointed out that the composition-dependent dust formation and non-gray radiative transfer might play a crucial role if one wants to model the dusty wind correctly. \citet{freytag2008} performed a 3D calculation of the AGB atmosphere with non-local radiation transfer in an explicit scheme. Their results not only exhibited the dust formation in the post-shock region but also showed the AGB winds with reasonable mass-loss rates. At the same time, radiation may affect the dynamics of the circumbinary disks in AGB binaries but is still poorly understood. A critical physics in the circumbinary disks is the evolution of the dust and the scattering of the radiation by the dust.

The \new{hydrodynamic problem with realistic} equation of state (EoS) has not been studied systematically in the context of the WRLOF. In AGB stars' atmospheres, the internal energy change in phase transitions, e.g., $\ce{H2}\rightleftharpoons\ce{H}$ and $\ce{H}\rightleftharpoons\ce{H+}+\ce{e-}$ is significant compared to the kinetic part of thermal energy \citep{chen2019} and radiation energy. \citet{freytag2008} used a Roe-type Riemann solver with a tabulated equation of state to resolve the ionization. Phase transitions such as the dissociation of \ce{H2} and the ionization of \ce{H} take place in the accreting flow, too \citep{omukai1998}. These phase transitions are very similar to the ones in star formation, and many similar physical phenomena may happen around the accreting MS stars in AGB binaries.

\new{The morphology of the outflow and the formation of accretion disk in evolved binary system has been studied by many colleagues \citep{theuns1993,mastrodemos1998,mastrodemos1999,devalborro2009,chen2017,devalborro2017,liu2017,saladino2018,saladino2019a,kim2019}, to name but a few. All the above mentioned works were successful in predicting the formation of spiral structure outflow. \citet{mastrodemos1999} was among the first to model the hydrodynamics of the circumstellar material which is caused by the focused wind in evolved binary. \citet[hereafter C17]{chen2017} performed a non-local radiation transfer calculation in 2D, and they observed the formation of circumbinary disks in some AGB binaries. Circumbinary disk does not appear in some recent works \citep{devalborro2017,liu2017,saladino2018,saladino2019a} where no such calculation is carried out. In this work, we resolve the radiation transfer in different azimuthal angles to improve the modelling of the dynamics of the circumbinary disks.}

Numerical accuracy is of paramount importance if someone tries to tackle a system that is highly nonlinear. Grid-based codes in an Eulerian picture and Smoothed-Particle Hydrodynamics (SPH) \citep{lucy1977,gingold1977} codes in a Lagrangian picture are two popular kinds of numerical codes that are suitable for radiation-hydrodynamic astrophysical problems. With mesh-refinement techniques and high order algorithms, Eulerian grid-based codes are effective in resolving shocks but are not Galilean invariant \citep{springel2010}. The angular momentum is in general not strictly conserved in grid-based codes. On the other hand, SPH codes need more effort in resolving strong shocks \citep{tasker2008} and instabilities \citep{agertz2007} while good SPH codes can evolve angular momentum conservatively. Both methods have been used in the study of the orbital evolution in AGB binaries. \citet{chen2018} studied the change of the binary separation by extracting the information of the mass transfer rate, mass-loss rate, and angular momentum loss rate in their Cartesian grid-based radiation hydrodynamic model \citepalias{chen2017}. The error in angular momentum conservation was tested and showed that it would not affect their conclusions in \citet{chen2018}. SPH method has been used to study the evolution of the binary separation \citep{saladino2019a} and eccentricity \citep{saladino2019b}. \newrr{We find that many previous numerical studies of AGB binaries with SPH codes do not model the pulsating AGB winds.} In our opinion, shocks are very important in AGB binaries. In the atmosphere of the AGB star, shocks can levitate the atmosphere to the radius where dust could condensate \citep{lamers1999,freytag2017}. Some post-shock material may fall back to the AGB star. Without shocks, the mass-loss rate of AGB winds is usually too low. Another important shock in an AGB binary is the bow shock near the secondary. The region within the bow shock is a domain of dependence of the accretion disk. Material that crosses this bow shock may be eventually accreted by the secondary.

In this research, we present a 3D radiation hydrodynamic model for the AGB binary systems \new{to illustrate the transition from the BHL accretion to the WRLOF}. We carried out Four simulations on the Cedar cluster of Compute Canada\footnote{https://www.computecanada.ca}. Each simulation costs about 20 core years, or equivalently, about two months on 128 cores. The rest of this paper is organized as follows: we discuss the adopted physics models in Section \ref{sec:physics}. Section \ref{sec:method} outlines the governing equations of our 3D radiation-hydrodynamic binary model. Section \ref{sec:setup} discusses the setup of the simulation domain. Section \ref{sec:boundary} describes the boundary conditions for the AGB star and secondary. The results of the single AGB star, mass transfer efficiency, outflow morphology, outflow angular distribution, and orbital stability can be found in Section \ref{sec:results}. We conclude our work and discuss the implications of our results in Section \ref{sec:conclusion}.

\section{Physics}\label{sec:physics}

\subsection{Overall setup of the problem}

We consider binary systems that consist of an AGB star with a fixed mass of 1.02 \msun\ and a MS star with a fixed mass of 0.51 \msun. The separation of the binary varies from 5.4 AU to 6.6 AU. This range of orbital separations was found to encapsulate the transition from the WRLOF to BHL accretion, as will be discussed in Section \ref{sec:morphology} and \ref{sec:beta}. The orbit of the binary is considered to be circular.

The stellar model of the primary star is obtained with detailed stellar evolution code and then the stellar model is used as a boundary condition (see \S\ref{sec:boundary}).
Specifically, we use {\tt MESA} (Modules of
Experiments in Stellar Astrophysics), release 10398 \citep{paxton2011,paxton2013,paxton2015,paxton2018}. 
We evolve a 1.5 \msun\  zero-age main-sequence (ZAMS) star with an initial $Z=0.02$ and $X=0.7$ until the star has reached its AGB phase. The same hydrogen content is used for the matter expelled from the AGB star.
We pick the model in the middle of the fast mass-loss phase  
when the mass of the star has decreased to 1.02 \msun. 
During the mass-loss, the star goes through lengthily 
episodes of expansions and contractions.
The specific model we chose for this work 
has an effective temperature $T_{\rm eff}=2874$ K, luminosity $L_{\rm AGB}=4384$ \lsun\ and radius of the photosphere $r_{\rm photo}=267$ \rsun.

\subsection{Dust formation and destruction}\label{sec:dustformation}

Dust forms \new{at} about 2-3 stellar radii of the AGB star \citep{hofner2018} and may be destroyed near the secondary (see below). We assume that destruction and formation of dust are dependent on the gas density $\rho$ and the gas temperature $T$, and dust does not exist in the following regions:
\begin{enumerate}
    \item $T>10^{4}$ K.
    \item $T_{\rm eq}>T_{\rm{sub}}$, where $T_{\rm{sub}}$ is the dust sublimation temperature.
    \item $\rho>10^{-15}$\ \gcmc\ and $T>1300$ K.
\end{enumerate}

Criterion 1 ensures that there is no dust in the very high-temperature region. In the simulations, strong shocks near the accretor may raise the temperature to $10^{4}$ K thus destroy the dust.

The physical meaning of criterion 2 is that the dust is removed if its "radiation equilibrium" temperature is higher than the sublimation temperature. By "radiation equilibrium" temperature we mean that a spherical dust particle is heated to the temperature that its blackbody radiation can balance the radiation it absorbs if the emission opacity and the absorption opacity are equal, i.e.,
\begin{equation}\label{eqn:balance}
    \int_{0}^{\infty}\kappa_{\lambda}I_{\lambda}\pi a_{\rm dust}^{2}d\lambda=\int_{0}^{\infty}\kappa_{\lambda}B_{\lambda}(T_{\rm eq})4\pi^{2}a_{\rm dust}^{2}d\lambda,
\end{equation}
where $\lambda,\ \kappa_{\lambda},\ I_{\lambda}$, $B_{\lambda}$ are the wavelength, Planck opacity, incoming radiation intensity, and Planck function, respectively. $a_{\rm dust}$ is the radius of the spherical dust grain. If $\kappa_{\lambda}$ is a constant, Equation \ref{eqn:balance} can be simplified to
\begin{equation}
    \int_{0}^{\infty}I_{\lambda}d\lambda=\int_{0}^{\infty}4\pi B_{\lambda}(T_{\rm eq})d\lambda.
\end{equation}
If only short wavelength radiation can be absorbed by the newly formed dust, the fraction of that absorbed radiation can be expressed by
\begin{equation}\label{eqn:alpha}
    \alpha=\frac{\int_{\lambda_{1}}^{\lambda_{2}}\pi B_{\lambda}(T_{\rm{eff}})d\lambda}{\sigma_{\rm{sb}}T_{\rm{eff}}^{4}} \ .
\end{equation}
Here $\sigma_{\rm{sb}}$ is the Stefan-Boltzmann constant. For $\lambda_1=0.01$ \micron\ and
$\lambda_2=0.91$ \micron, 
$\alpha=0.1875$. This range of wavelengths likely covers the majority of the photons that could be effectively absorbed by the newly formed small ($a_{\rm{dust}}<0.1\ \mu$m) dust grains.

If the material between the AGB star and the dust formation region is transparent, $I_{\lambda}=B_{\lambda}(T_{\rm{eff}})$. Make use of Equation \ref{eqn:balance} and \ref{eqn:alpha} and further assume that $\kappa_{\lambda}$ is a constant, we can find 
\begin{equation}
T_{\rm eq}=\left(\frac{\alpha L_{\rm{p}}}{4\pi\sigma_{\rm{sb}} r_{\rm{p}}^{2}}\right)^{0.25},
\end{equation}
where $r_{p}$ is the distance from the dust to the center of the star. A dust formation radius can be calculated if a sublimation temperature $T_{\rm sub}$ is given
\begin{equation}
    r_{\rm{p}}=\frac{\sqrt{\alpha}T_{\rm{eff}}^{2}r_{\rm{photo}}}{T_{\rm{sub}}^{2}}.
\end{equation}
When $T_{\rm sub}=1300$ K, $r_{\rm{dust}}\approx2.12r_{\rm photo}$.

\begin{table}[]
    \centering
    \begin{tabular}{ccc}\hline
    composition &   $r_{\rm pyro}/r_{\rm photo}$    &   $r_{\rm pyro}/r_{\rm photo}$\\\hline
    \ce{Mg_{0.5}Fe_{0.5}SiO3}    &  2.85    &   2.66    \\
    \ce{Mg_{0.6}Fe_{0.4}SiO3}    &  3.00    &   2.79    \\
    \ce{Mg_{0.7}Fe_{0.3}SiO3}    &  1.71    &   1.63    \\
    \ce{Mg_{0.8}Fe_{0.2}SiO3}    &  1.43    &   1.38    \\\hline
    \end{tabular}
    \caption{The second and third column show the formation radius of the test dust grain whose radius is $a_{\rm dust}=0.0535$\ \micron\ and $a_{\rm dust}=0.01$\ \micron, respectively.}
    \label{tab:formationradii}
\end{table}

\begin{figure}
    \centering
    \includegraphics[width=\columnwidth]{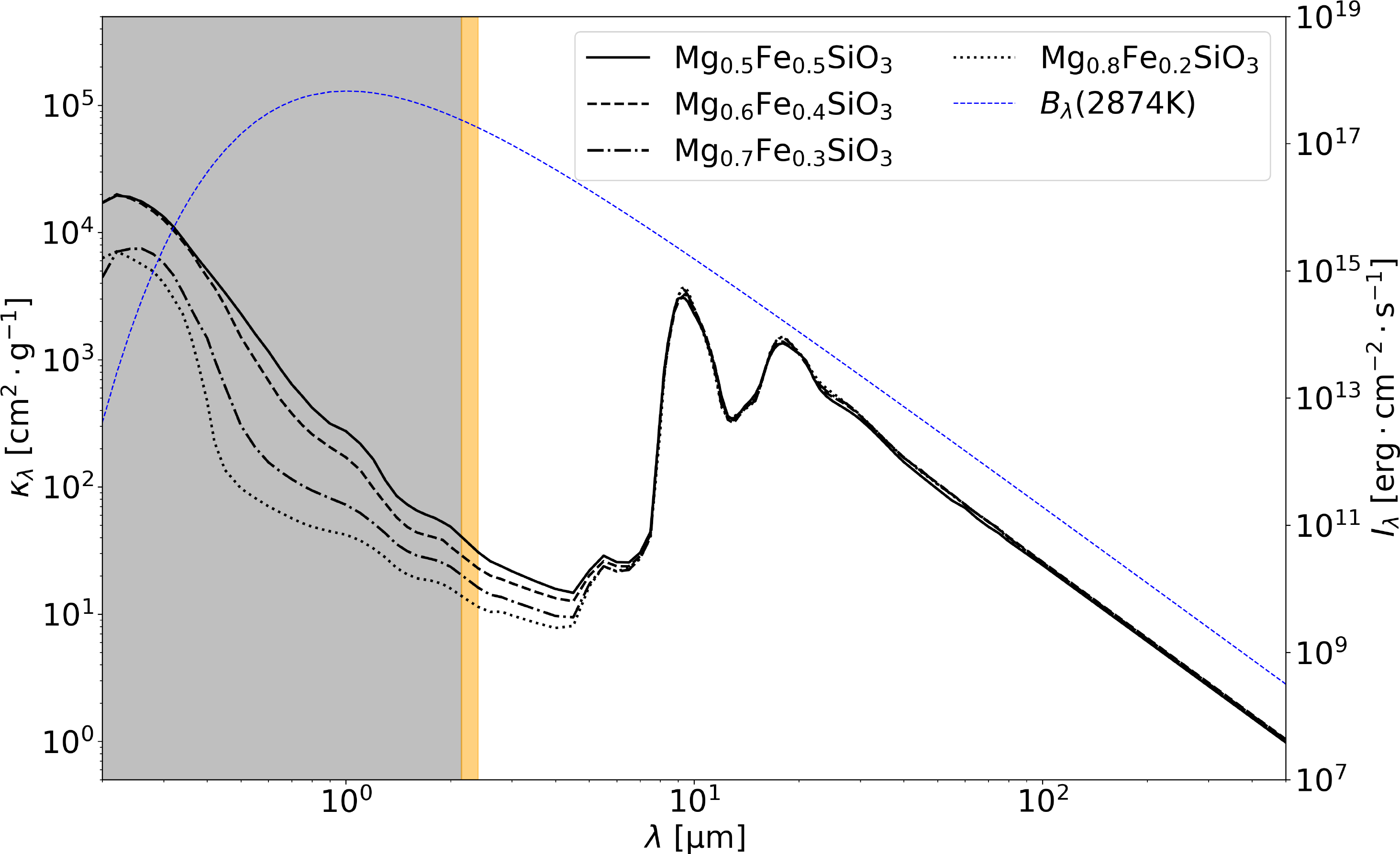}
    \caption{Examples of the wavelength dependence of the absorption opacity of dust. The left y-axis is the opacity and the right y-axis is the radiation power. The black lines show the opacity of different dust species. The blue dashed line shows the black-body spectrum for $T_{\rm{eff}}=2874$ K. Gray band covers the wavelengths $0.01\ \mu\rm{m}\le\lambda\le2.15\ \mu\rm{m}$ and the yellow region covers the wavelengths $2.15\ \mu\rm{m}\le\lambda\le2.4\ \mu\rm{m}$. The shown opacities are found assuming the dust's number density distribution of $dn_{\rm{dust}}\approx a_{\rm{dust}}^{-3.5}$ for dust radius $0.1\ \mu\rm{m}<a_{\rm{dust}}<0.3\ \mu\rm{m}$ \citep{mathis1977}.}  
    \label{fig:opacity}
\end{figure}

Indeed, $\kappa_{\lambda}$ for dust is a function that has a greater value in the short wavelength, see Figure \ref{fig:opacity} where we show $\kappa_\lambda$ for pyroxene with different concentration of iron. We take pyroxene as an example because it has a high sublimation temperature of about $1500$ K \citep{kobayashi2011} and exists near M-type AGB stars \citep{bladh2013,bladh2015}. We use databases of Dust Optical Properties \footnote{https://www.astro.uni-jena.de/Laboratory/OCDB/index.html} for the optical properties of the specific pyroxene composition \citep{jaeger1994,dorschner1995}. The dust opacity is calculated using the Mie model \citep{matzler2002}.  We use the variable $\kappa_{\lambda}$ and the whole spectrum to find the formation radii for several types of dust grains, see Table \ref{tab:formationradii}. We can see that the radius predicted by our simplified dust formation model falls in the range of the formation radius of pyroxene.

In our simulations, $r_{\rm dust}$ is found using the actual incoming intensity. The incoming intensity may be attenuated by the matter between the AGB star and the dust formation region, i.e., it depends on the dust-free optical depth $\tau$, which itself is the function of the direction from the AGB star. As a result, the location of dust formation or destruction depends on the direction from the AGB star and is
\begin{equation}
    r_{\rm dust}(\theta,\phi)=\frac{\sqrt{\alpha}T_{\rm{eff}}^{2}r_{\rm{photo}}e^{-\tau(\theta,\phi)}}{T_{\rm{sub}}^{2}},
\end{equation}
where $\theta$ is the polar angle and $\phi$ is the azimuthal angle in the spherical coordinate system centered at the AGB star's center (see Appendix \ref{sec:tau} for detail).

The purpose of criterion 3 is to describe dust destruction around the secondary star. As the density gets higher near the secondary, the dust and gas may come into LTE again and the dust sublimates.

Dust formation is a complicated subject because many species form at different $T$s, and the formation sequence is of great importance \citep{jeong2003,woitke2006a}. We anticipate that our approach is simplified. We note however that our goal is not to study the dust formation but to use a relatively reasonable dust formation model to drive a phenomenologically pulsating AGB wind.

\subsection{Dust and its ability to capture radiation}\label{sec:f}

We integrate the blackbody spectrum for $T_{\rm eff}=2874$ K and find that within the band $0.01\ \mu\rm{m}\le\lambda\le2.15\ \mu\rm{m}$ it covers $75\%$ of the total radiation power; the region $2.15\ \mu\rm{m}\le\lambda\le2.4\ \mu\rm{m}$ covers an additional $5\%$. In this work, we focus on the absorption of small dust grains. In the case that the dust may grow to 1\micron, the reddening of the spectrum may cover the whole K band \citep{bladh2013,bladh2015}. In summary, the overall momentum contribution of the blackbody radiation beyond the K band is relatively small, as a result, we only take $75\%$ of the incoming radiation energy as the momentum budget that can be transferred to the dust; we hence choose to use
\begin{equation}
    L_{\rm eff}=fL_{\rm AGB}=0.75L_{\rm AGB}
\end{equation}
as the effective luminosity for the momentum transfer.

\subsection{Dust opacity}\label{sec:opacity}

When $r_{\rm{dust}}$ is known, we use an opacity profile that mimics the dust growth from the dust-free region to the dusty region
\begin{equation}\label{eqn:opacity}
    \kappa=\frac{\kappa_{\rm{dust}}}{1+\exp{-2\frac{r_{\rm{p}}-r_{\rm_{dust}}-r_{\rm{scale}}}{r_{\rm{scale}}}}}+\kappa_{\rm{mol}}\ .
\end{equation}
Here $\kappa_{\rm{mol}}=2.5\times10^{-4}$ \cmsg\ is the dust-free opacity and $\kappa_{\rm{dust}}=6$ \cmsg. $r_{\rm{scale}}=0.16$ AU is the adopted length-scale which mimics the dust growth length-scale. We introduce the corresponding radius where the opacity of the dust-gas mixture reaches the half of its maximum
\begin{equation}
    r_{\rm{half}}=r_{\rm_{dust}}+r_{\rm{scale}}\approx2.75\ \rm{AU} \ .
\end{equation}

The choice of $r_{\rm{scale}}$ is empirical but an order of magnitude estimation could be made to justify the choice. In our model, the characteristic speed of a pulsation at the dust formation radius is $v_{\rm{ch}}\approx5$ \kms. We can estimate the implied timescale for the dust growth as
\begin{equation}
    t_{\rm{growth}}\approx2r_{\rm{scale}}/v_{\rm{ch}}\approx0.3\ \rm{year}.
\end{equation}
The multiple 2 comes from \new{the adopted opacity profile whose} opacity grows to half of its maximum in about $r_{\rm{scale}}$ distance. This value is smaller than the typical dynamical timescale and pulsation period ($\sim$1 year) of an AGB star. The AGB stars exhibit variation in the optical light-curves with periods of 200-1000 days which implies that the dust formation time should be smaller than the period of the light-curves.


\subsection{Optically thin cooling}\label{sec:cooling}

We use the following two groups of processes for the cooling process.
\begin{enumerate}
    \item The rotational and vibrational cooling of \ce{H2}, \ce{H2O}, and \ce{CO} \citep{neufeld1993}.
    \item Cooling due to the collisional ionization, recombination, collisional excitation, and Bremsstralung of \ce{H} and \ce{H+} \citep{cen1992}.
\end{enumerate}
We set \citep{neufeld1993,lacy1994}
\begin{eqnarray}
    n_{\ce{H2O}}/n_{\ce{H2}}&=&2\times10^{-4},  \\
    n_{\ce{CO}}/n_{\ce{H2}}&=&2\times10^{-4}.
\end{eqnarray}
In principle, the abundance of molecules cannot be known without carrying out chemical reaction calculations. In the AGB wind, the C/O ratio is vital to the species of the molecules and dust \citep{hofner2018}. The ratios we adopted in our simulations make sense only for some oxygen-rich AGB stars.

We calculate $n_{\ce{H+}},n_{\ce{H}}$, and $n_{\ce{H2}}$ by solving the thermal equilibrium problem, i.e., the Saha equations with the local $\rho$ and $T$ \citep{chen2019}
\begin{eqnarray}
    \ce{H}&\rightleftharpoons&\ce{H+}+\ce{e-},  \\
    \ce{H2}&\rightleftharpoons&2\ce{H}.
\end{eqnarray}

In summary, the cooling strength is
\begin{equation}\label{eqn:cool}
    \dot{\Lambda}=\dot{\Lambda}_\text{\ce{H2}}+\dot{\Lambda}_\text{\ce{H2O}}+\dot{\Lambda}_\text{\ce{CO}}+\dot{\Lambda}_\text{\ce{H},\ce{H+}}.
\end{equation}

Ideally, optically thin cooling should only be applied to the optically thin region. In this research, we use $\rho$ alone to identify whether an individual cell is optically thick. If $\rho>10^{-9}$ \gcmc, we assume such a cell is optically thick and the optically thin cooling will be turned off. However, with such a criterion, the vicinity of the accretion disk may become optically thick. The cooling will be halted and the temperature may rise rapidly to more than $10^{5}$ K. To make the temperature profile smoother and keep the accretion process efficient, we keep the cooling mechanism around the secondary (inside the accretion disk). In a real accretion disk, radiation (thermal and non-thermal) should be able to carry away energy. We currently do not have such a consistent method to let the energy goes away, therefore, we keep the simple cooling method. This may lead to an overestimate of the cooling strength inside the accretion disk.

\section{Governing equations}\label{sec:method}

The governing equations are the compressible Euler equations in a rotating frame with relevant sink and source terms. We solve the radiation hydrodynamic equations by {\tt AstroBEAR} with static mesh refinement \citep{carroll2013}.
\begin{eqnarray}\label{eqn:governning1}
    \frac{\partial\rho}{\partial t}+\nabla\cdot(\rho\Vec{\mathbf{v}})&=&0,\\\label{eqn:governning2}
    \frac{\partial\rho\Vec{\mathbf{v}}}{\partial t}+\nabla\cdot(\rho\Vec{\mathbf{v}}\Vec{\mathbf{v}})&=&-\nabla p+\rho(\Vec{\mathbf{a}}_{\rm{rad}}+\Vec{\mathbf{a}}_{\rm{g}}+\Vec{\mathbf{a}}_{\rm{i}}),\\\label{eqn:governning3}
    \frac{\partial E}{\partial t}+\nabla\cdot[\Vec{\mathbf{v}}(E+p)]&=&\dot{\Lambda}+\rho\Vec{\mathbf{v}}\cdot(\Vec{\mathbf{a}}_{\rm{rad}}+\Vec{\mathbf{a}}_{\rm{g}}+\Vec{\mathbf{a}}_{\rm{i}}),
\end{eqnarray}
where $t,\rho,p$, and $E$ are the time, density, pressure, and total energy, respectively. $\Vec{\mathbf{v}}$ the is the velocity in the rotating frame. $\Vec{\mathbf{a}}_{\rm{rad}},\ \Vec{\mathbf{a}}_{\rm{g}},\ \Vec{\mathbf{a}}_{\rm{i}}$, and $\dot{\Lambda}$ are the acceleration due to the radiative momentum transfer, gravity of the two stars, and inertial force, and the energy change rate due to the optically thin cooling. We will discuss the considered physics in detail in the following subsections. Also, we adopt a perfect gas EoS,
\begin{eqnarray}
    p&=&\frac{\rho k_{\rm{b}}T}{\mu m_{\rm{amu}}},\\
    \epsilon&=&\frac{p}{(\gamma-1)\rho},\\
    E&=&\rho(\epsilon+\frac{v^{2}}{2}),
\end{eqnarray}
where $k_{\rm{b}}$ and $m_{\rm{amu}}$ are the Boltzmann constant and atomic mass unit. The mean atomic weight is $\mu=1.3$ and the adiabatic index is $\gamma=5/3$. We solve the hydrodynamics by the Harten-Lax-van Leer-Contact (HLLC) Riemann solver with the corner transport upwind method \citep{colella1990} and piece-wise linear reconstruction.

\subsection{Gravitational force and inertial force}\label{sec:gravity}

$\Vec{\mathbf{a}}_{\rm{g}}$ is the gravitational acceleration
\begin{equation}
    \Vec{\mathbf{a}}_\text{g}=-\frac{Gm_\text{p}\Hat{\mathbf{r}}_\text{p}}{r_\text{p}^{2}}-\frac{Gm_\text{s}\Hat{\mathbf{r}}_\text{s}}{r_\text{s}^{2}},
\end{equation}
where $G$ is the gravitational constant and 
\begin{eqnarray}
    \Vec{\mathbf{r}}_{\rm{p}}&=&\Vec{\mathbf{r}}-\Vec{\mathbf{l}}_{\rm{p}},   \\
    \Vec{\mathbf{r}}_{\rm{s}}&=&\Vec{\mathbf{r}}-\Vec{\mathbf{l}}_{\rm{s}}.
\end{eqnarray}
Subscript p and s stand for the primary star and the secondary star, respectively. $\Vec{\mathbf{l}}_{\rm{p}}$, $\Vec{\mathbf{l}}_{\rm{s}}$, and $\Vec{\mathbf{r}}$ are the Cartesian coordinates of the primary, the secondary, and the cell. We choose the origin to be at the center of mass of the binary system. Self-gravity is ignored because the total mass of the ejected material in all of our simulations is less than $5\times10^{-5}$\ \msun\ which is small compared to the mass of the binary stars. The outflows also have some symmetry which may cancel out most of the net effect of self-gravity.

$\Vec{\mathbf{a}}_{\rm{i}}$ is the acceleration due to the inertial force in the rotating frame.
\begin{equation}
    \Vec{\mathbf{a}}_\text{i}=-2\Vec{\mathbf{\Omega}}_\text{b}\times \Vec{\mathbf{v}}-\Vec{\mathbf{\Omega}}_\text{b}\times(\Vec{\mathbf{\Omega}}_\text{b}\times\Vec{\mathbf{r}}),
\end{equation}
where $\Vec{\mathbf{\Omega}}_{\rm{b}}=(0,0,\omega)$ is the pseudo-vector of the rotating frame, and
\begin{equation}
    \omega=\sqrt{G(m_{\rm p}+m_{\rm s})/d^3}
\end{equation}
is the orbital frequency. $d$ is the binary separation. Throughout our simulations, $\Vec{\mathbf{l}}_{\rm{p}}$, $\Vec{\mathbf{l}}_{\rm{s}}$, and $\Vec{\mathbf{\Omega}}_{\rm{b}}$ are kept constant for computational simplicity. That means we do not track the orbital evolution of the binaries in our simulations. We examine this assumption in Section \ref{sec:orbit}.

\subsection{Non-local momentum transfer by radiation}\label{sec:radforce}

In this work, we only consider a single interaction between the photon and dust, i.e., we adopt that the photons are destroyed by the dust once they interact with the dust. This is a reasonable approach as the dust temperature is low, the thermal radiation from a dust grain is less likely to be absorbed by another dust grain. On the other hand, scattering by the dust usually only changes the photon's direction by a small angle \citep{henyey1941} thus the anisotropy of the radiation from the AGB star may not be strong. Therefore, we only consider the radiation in the radial direction of the AGB star.

$\Vec{\mathbf{a}}_{\rm{rad}}$ is the acceleration due to the radiative pressure from the primary star. The radiation force is given by,
\begin{equation}\label{eqn:arad}
    \Vec{\mathbf{a}}_{\rm{rad}}=\frac{\kappa L_{\rm eff}e^{-\tau}\Hat{\mathbf{r}}_{\rm p}}{4\pi c r_{\rm p}^{2}}.
\end{equation}
Here $c$ is the speed of light. $\tau(x,y,z)$ is obtained by linearly interpolating $\tau(r,\theta,\phi)$. $\kappa$ is given by Equation \ref{eqn:opacity}. We treat the AGB star as a point radiation source.

\subsection{Time-step and control of the integration}

Some of the physics we used, e.g., the cooling, may operate on the timescale shorter than the hydro time-step in a few cells. This makes the energy equation stiff. We solve the stiff ODE problem by comparing the 4th and 5th order Runge-Kutta solution and adaptively evolve the time-step \citep{press1992}. For each integration step, we only allow a maximum of $20\%$ change in the internal energy, i.e.,
\begin{equation}
    \epsilon^{\rm{n}+1}>0.8\epsilon^{\rm{n}},
\end{equation}
here n is the step number of sub-cycle of the integration.

\section{The simulation domain}\label{sec:setup}

We fix the initial mass ($m_{\rm{p}}=1.02$ \msun) of the AGB star and the secondary  ($m_{\rm{s}}=0.51$ \msun) in all of the models. We consider four binary separations, $d=5.4,5.7,6.0,6.6$ AU.

The simulation box is the same for all the models. It is $48\ \rm{AU}\times48\ \rm{AU}\times24\ \rm{AU}$ in the x, y, and z directions, respectively. The base resolution of the domain is $60\times60\times30$ cells and we use five levels of mesh refinement. Each level of mesh refinement doubles the resolution in each dimension. The physical size of the finest cell is $(2.5\times10^{-2}\ \rm{AU})^{3}$.

The actual simulation domain is a cylinder whose axis coincides with the z-axis. The radius and the height of the cylinder are $r_{\rm{domain}}=23.5$ AU and $h_{\rm{domain}}=24$ AU, respectively. This cylinder is inside the simulation box. Radially outward supersonic flow in the lab frame is set outside the cylinder. The temperature of the supersonic flow is $1000$ K and the speed is $15$ \kms. Such a supersonic flow ensures that there is no information propagating into the cylinder from its side surface. The boundary condition of the top and bottom surfaces is set diode, i.e, only outgoing flow is allowed.

The refined region around the secondary is two con-axis cylinders, one with five levels of refinement has a height of $0.5$ AU and a radius of $1.35$ AU. The second cylinder has four levels of refinement and has a height of $0.7$ AU and a radius of $2.7$ AU. We chose such configuration so that the bigger cylinder covers the circularization radius \citep{frank2002}, and the smaller cylinder encapsulates the accretion disk. The AGB star is resolved by a sphere with four levels of refinement and the radius of the sphere is $2.85$ AU. The sphere is aimed to resolve the dust forming region. We resolve the equatorial zone by using a cylinder with three levels of mesh refinement. The height and radius of the equator's cylinder are $1.4$ AU and $22.5$ AU, respectively.

\section{Inner boundary conditions}\label{sec:boundary}

The primary star and the secondary star are two important inner boundaries in our simulations. We discuss their implementations separately.

\subsection{Primary star}\label{sec:primary}

\begin{figure}
    \centering
    \includegraphics[width=\columnwidth]{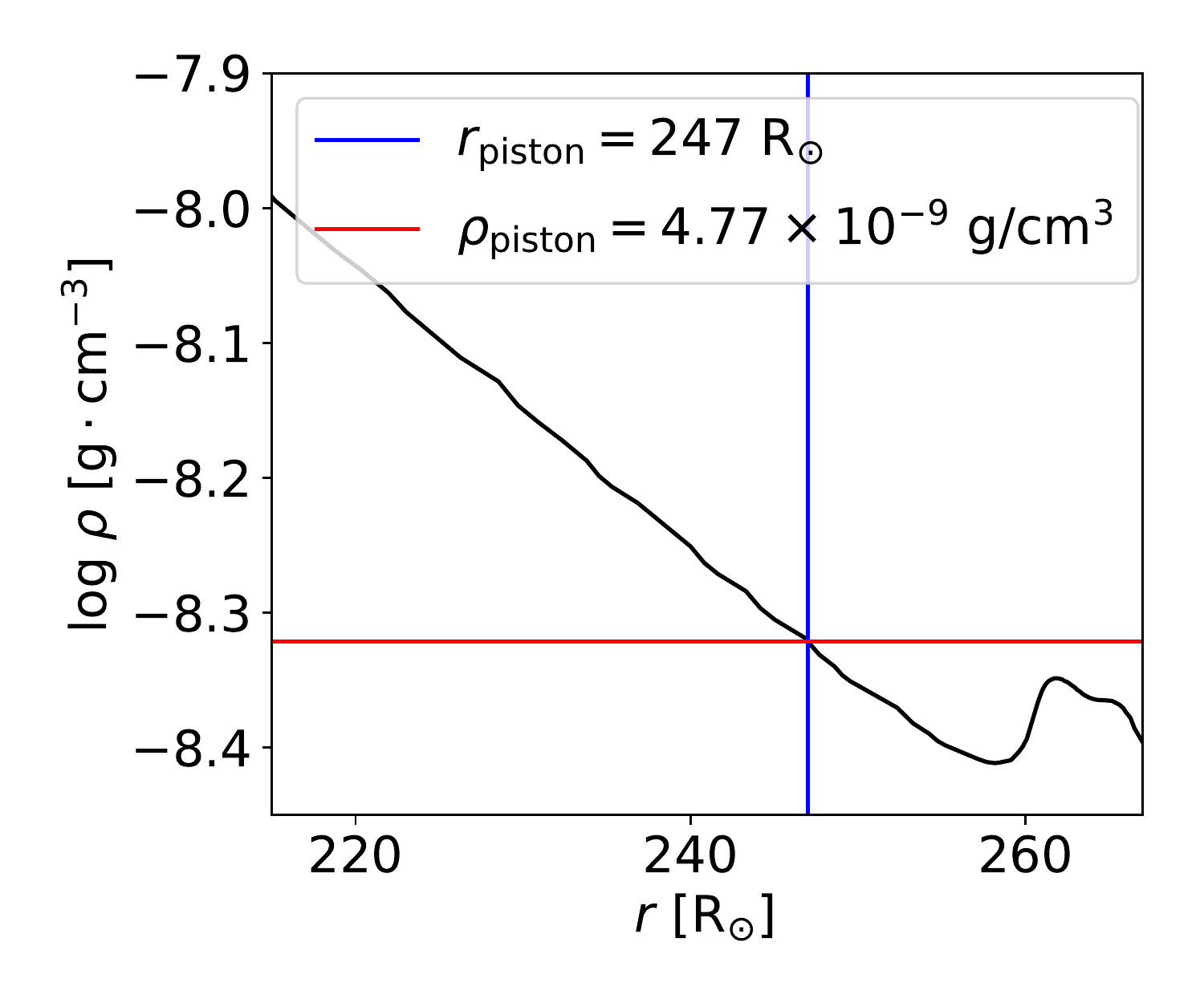}
    \includegraphics[width=\columnwidth]{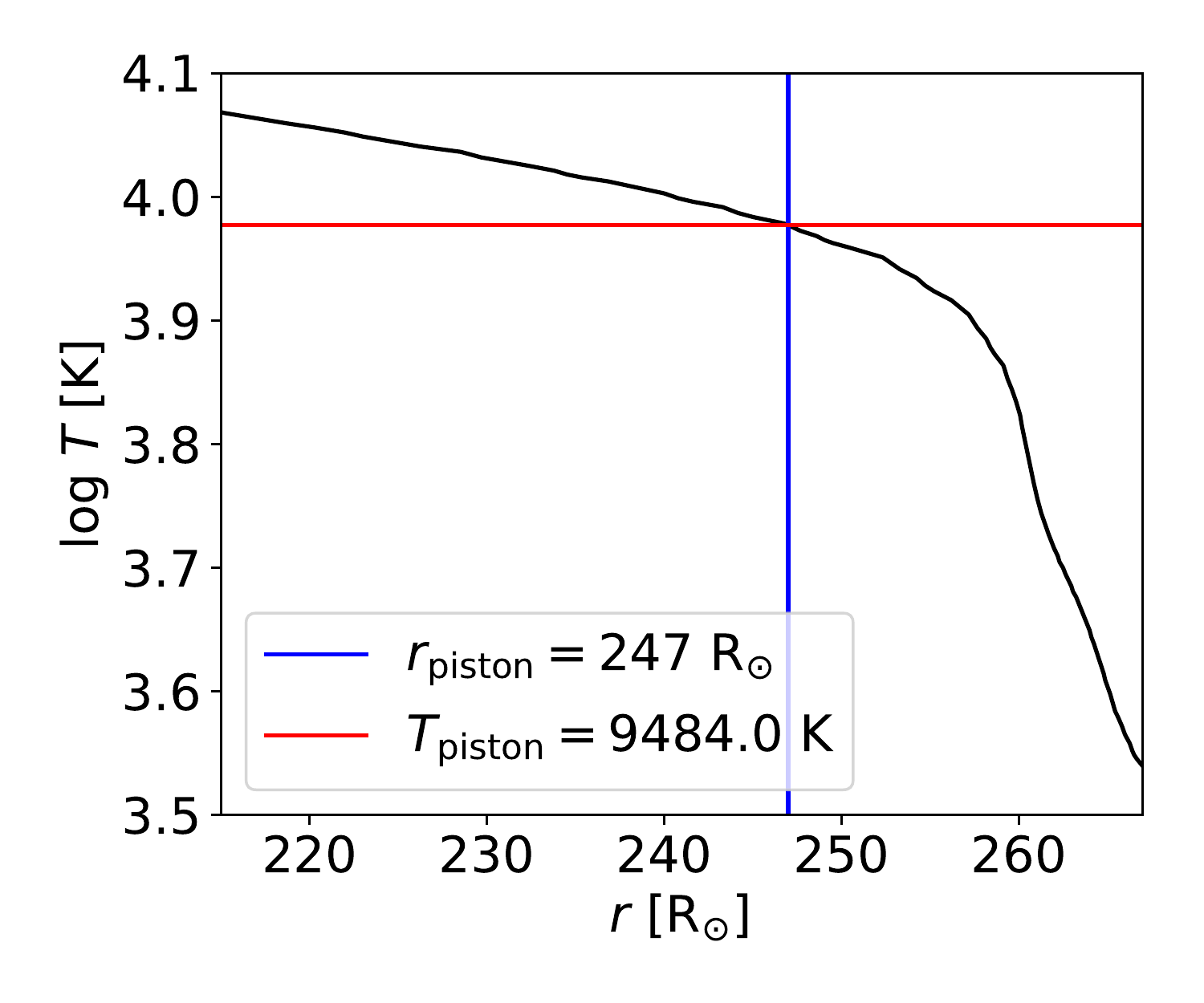}
    \caption{The zoomed in density and temperature profiles of the primary star produced by {\tt MESA}. The piston position is placed at $r_{\rm piston}=247$\ \rsun\ \new{from the center of the AGB star}.}
    \label{fig:mesa}
\end{figure}

The primary star is the AGB star. We use the piston model \citep{bowen1988} to approximate the pulsating AGB star. The boundary condition of the primary star determines the temperature, density, radius, velocity, and period of the piston. Figure \ref{fig:mesa} shows the density and temperature profiles of the AGB star as obtained from {\tt MESA}.

Similar to some other researches, we place the piston below the photosphere at $r_{\rm piston}=247$ \rsun\ \citep{bowen1988,liljegren2016}. At this radius, {\tt MESA} model has $T=9484$ K and $\rho=4.77\times10^{-9}$ \gcmc. The profiles and location of the piston determine the temperature and the density of the piston, which are kept constant throughout the simulations. The sound speed of the piston is roughly $10$ \kms. \new{To create a pulsating atmosphere}, we adopt a subsonic piston model. The radial velocity of the fluid inside the AGB star is described by
\begin{equation}\label{eqn:piston}
    v_{\rm{piston}}(r_{\rm{p}},t)=\frac{r_{\rm{p}}v_{\rm{amp}}}{r_{\rm piston}}\sin\left(\frac{2\pi t}{P_{\rm{piston}}}\right)\quad r_{\rm{p}}\le r_{\rm piston},
\end{equation}
where $v_{\rm{amp}}=9$ \kms\ is the amplitude of the piston (similar amplitude has also been used in \citet{jeong2003}) and $P_{\rm{piston}}=1$ year is the adopted period of the piston. The $v_{\rm{amp}}$ we choose corresponds to $0.9\ \textit{Mach}$. We have tested smaller amplitude and find that the subsequent mass-loss of the AGB star is lower. In this work, we adopt the AGB star with high mass-loss rate therefore we set the amplitude as 9 \kms. \new{We summarize the parameters of the AGB star model in Table \ref{tab:agb}}

\begin{table}[]
    \centering
    \begin{tabular}{cc}\hline
    parameter name  &  physical quantity  \\\hline
    $r_{\rm photo}$ &  $267$\ \rsun  \\
    $T_{\rm eff}$ &  2874 K   \\
    $L_{\rm AGB}$  &  4384\ \lsun  \\
    $L_{\rm eff}$  &  3288\ \lsun  \\
    $r_{\rm piston}$  &  $247$\ \rsun  \\
    $\rho_{\rm piston}$  &  $4.77\times10^{-9}$\ \gcmc  \\
    $T_{\rm piston}$  &  9484 K  \\
    $P_{\rm{piston}}$  &  1 year  \\
    $v_{\rm{amp}}$  &   9\ \kms \\\hline
    \end{tabular}
    \caption{Parameters of the AGB star model.}
    \label{tab:agb}
\end{table}

The AGB stars in all four binary models are not spinning in the lab frame. That means the AGB star is counter-rotating in the co-rotating frame. The reason for setting non-spinning AGB star is twofold:
\begin{enumerate}
    \item Complete tidal spin-orbit coupling state of an AGB star with its orbital rotation is not warranted, and the degree of that coupling is unknown. \citet{saladino2019a} analyzed the spin-orbit coupling in AGB binaries with high mass loss rates. From Figure D.1a of their paper, we can see that for $d\in[5-6]$AU and mass ratio $q=m_{\rm{p}}/m_{\rm{s}}=2$, the AGB star is spinning at $80\%-90\%$ of the orbital angular frequency. The AGB star they analyzed has a radius of 330 \rsun and is $33\%$ larger than the boundary of the AGB star (247 \rsun) we use. If the tidal spin-orbit coupling timescale scales with $r_{\rm p}/d$, we anticipate the spinning rate of our AGB star should be similar to their $d=8$ AU model which is about $60\%$.
    \item The single AGB star model we tested is a non-spinning one. We would have to test and verify many more single star models to justify the binary simulations with a spinning AGB star.
\end{enumerate}

By setting the spin of the AGB star to be zero, we are underestimating the initial angular momentum in the outflow. We now show that the underestimation is not big compared to the orbital angular momentum. The z-component of the specific angular momentum in a thin shell of a sphere can be calculated by 
\begin{equation}
    j_{\rm{z}}=\frac{2}{3}\omega r_{\rm piston}^{2},
\end{equation}
where $r_{\rm piston}=247$ \rsun\ is the radius of the piston layer. The orbital specific angular momentum at $d/3$ (the distance of the AGB star's center to the center of mass) is
\begin{equation}
    j_{\rm{p}}=\Omega_{\rm{b}}d^{2}/9.
\end{equation}
In our four binary simulations, the maximum ratio of $j_{\rm{z}}/j_{\rm{p}}$ is $0.27$. In a not fully co-rotating case, the ratio is smaller.

\subsection{Secondary star}\label{sec:secondary}

The secondary star is modeled as a point gravitational source. While it does not interact with the gas hydrodynamically, it can remove material in its vicinity that has become bound to it. The mass of the removed material is added to the point particle representing the secondary \citep{krumholz2004}. The point particle does not have the energy or spin angular momentum, therefore, the energy and angular momentum in the removed material is simply removed from the simulation. We define the 'vicinity' of the secondary to be $r_{\rm{acc}}=0.2$ AU or 8 finest cells in all of our simulations. We use spline softening \citep{federath2010} to soft the gravitational force of the secondary within $r_{\rm soft}=0.1$ AU.

\new{We turn off the accretion algorithm for the first 5 years in all the simulations to let the density around the secondary increase and the flow pattern form, then we turn on the accretion algorithm. We find this is helpful for the computational speed because the initial density contrast is very big and high speed flow emerges if we turn on the accretion algorithm. In our current model, the AGB binary problem is mainly a boundary value problem. Therefore, we think it is safe to turn off the accretion for a period of time as long as the simulation approaches a state that does not change drastically.}

\section{Results}\label{sec:results}

\subsection{Single star}\label{sec:single}

\begin{figure}
    \centering
    \includegraphics[width=\columnwidth]{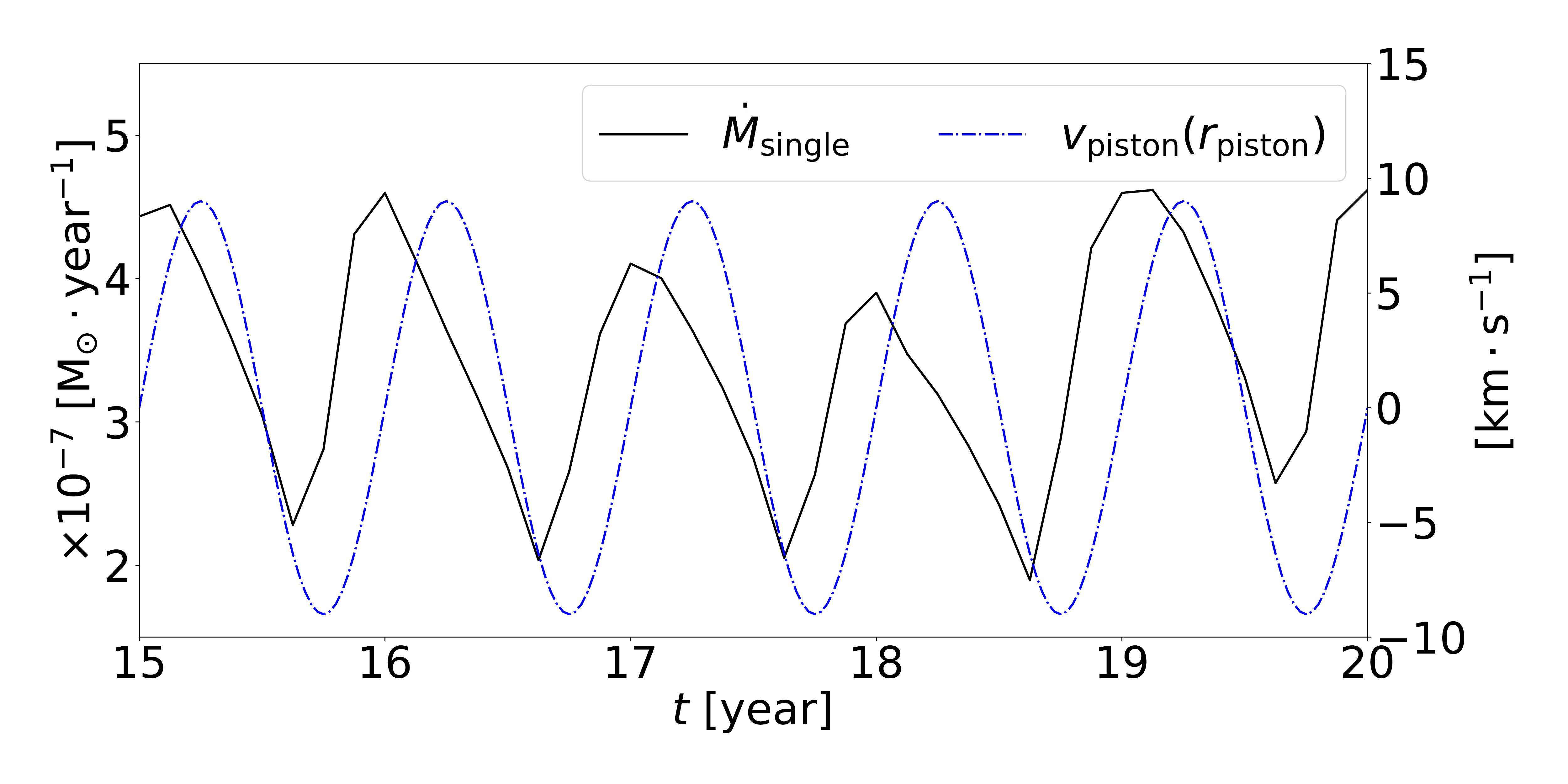}
    \includegraphics[width=\columnwidth]{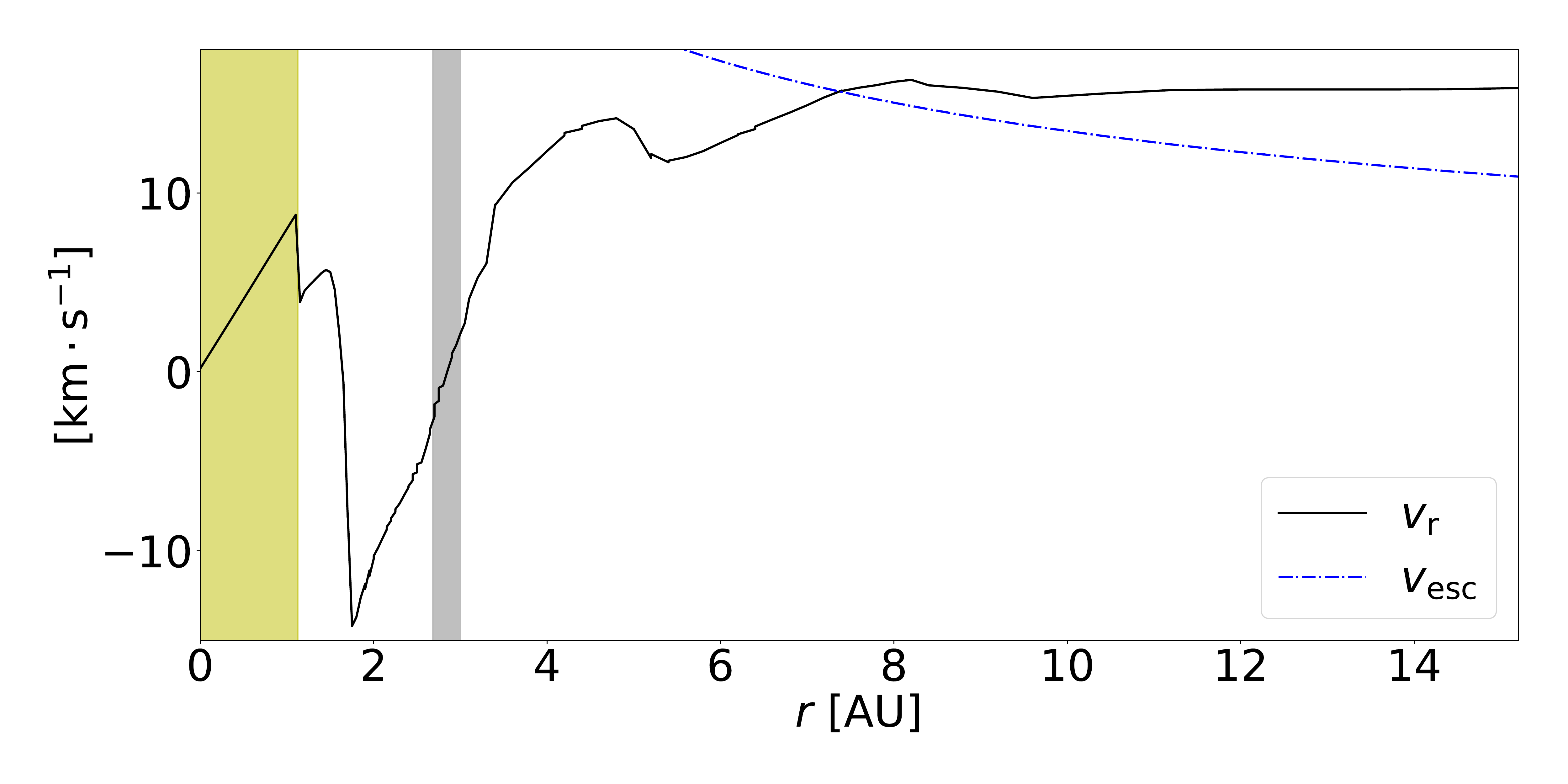}
    \includegraphics[width=\columnwidth]{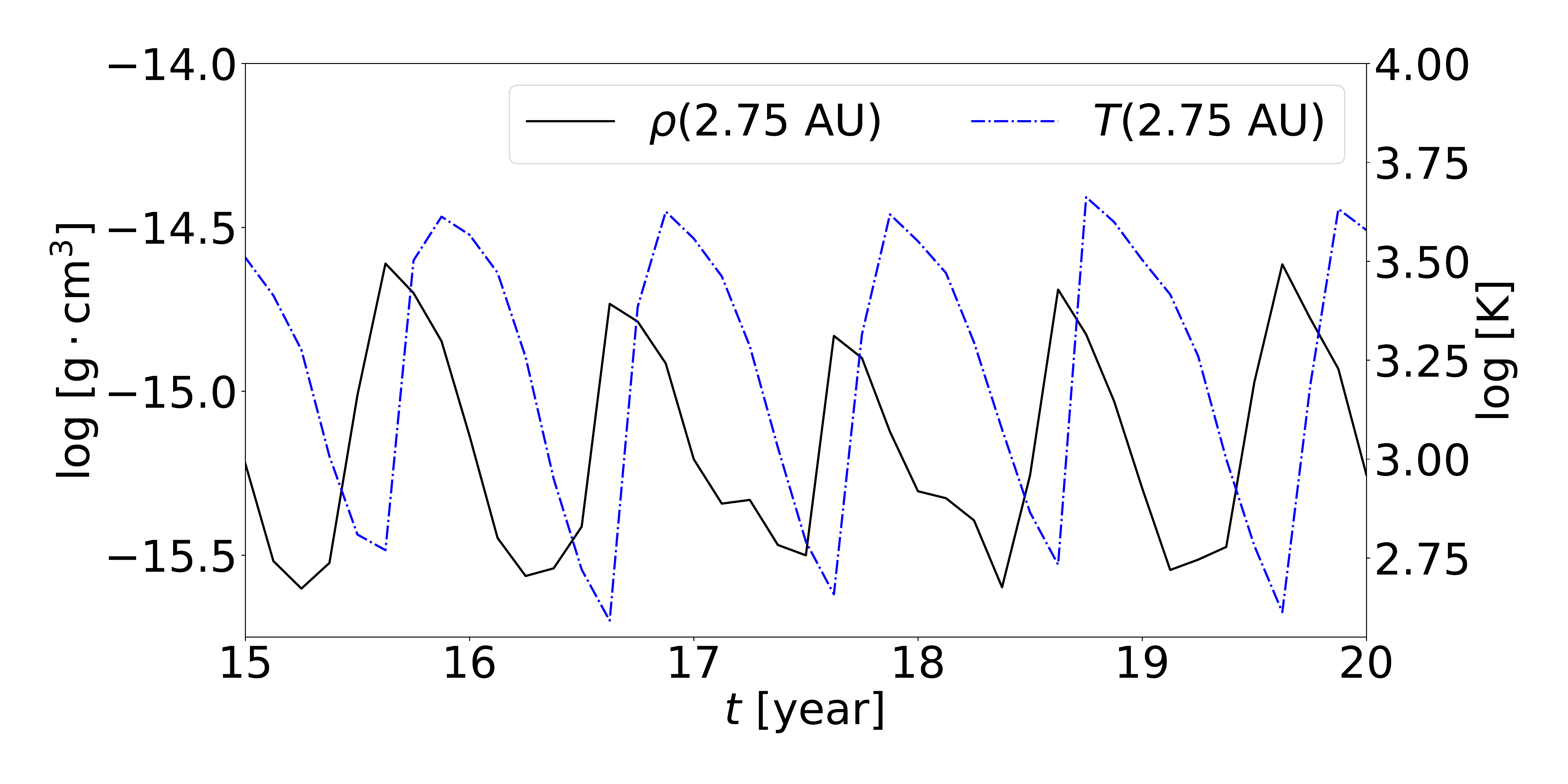}
    \caption{Top: the mass-loss rate of the single AGB star model and the radial velocity of the piston at its edge. Middle: the radial velocity profile of the wind of the x-axis and the escape speed. The AGB star is covered with the yellow band and the dust forming region is highlighted with the gray band. Bottom: the time-dependent gas density and gas temperature at $r_{\rm half}=2.75$ AU on the x-axis.}
    \label{fig:vrprofile}
\end{figure}

We first present the results of the single AGB star with the adopted physics and boundary conditions. Figure \ref{fig:vrprofile} shows the time-dependent mass-loss rate and piston's radial velocity, the wind's radial velocity profile and escape velocity and the time-dependent $\rho$ and $T$ measured at a distance $r=2.75$ AU from the center of the single AGB star. The phase difference between the mass-loss rate and the piston's radial velocity is irrelevant because it simply depends on the sampling shell of the mass flux.

We find that the mass-loss rate varies along with the phase of pulsation. We evaluate the mass flux at $r_{\rm{p}}=10$ AU. The average mass-loss rate is $\dot{M}_{\rm{single}}=3.45\times10^{-7}$ \msunyear. Due to falling back material and continuing pulsations, the gas between the dust formation region and the piston boundary is constantly shocked, which can be seen in the middle panel. The wind's velocity profile approaches a constant value at large distances. This terminal velocity of the AGB wind is $v_{\infty}\approx15.8$ \kms. A good summary of the relation of the luminosity, effective temperature, and mass-loss rate versus the wind terminal speed of M-type AGB stars can be found in Figure 18 in \citet{hofner2018}. The mass-loss rate in our model is within the range of the observed mass-loss rate of $10^{-7}-10^{-5}$ \msunyear. $v_{\infty}$ maybe a bit high. According to \citet{hofner2018}, a more reasonable wind terminal speed should be between $7$ \kms and $13$ \kms. However, this fact would not adversely affect the conclusions of this work as an AGB binary with a slow AGB wind is more likely to form a circumbinary disk (see Section \ref{sec:morphology}).

Phase differences of different quantities may play important roles in the mass-loss rate. \citet{liljegren2016} studied the impact of the phase difference between the pulsation and luminosity. They find that different phase shifts could lead to drastically different mass-loss rates. In our model, there is no phase difference in $\rho$ and $T$ in the piston. However, the phase between $\rho(2.75\ \rm{AU})$ and $T(2.75\ \rm{AU})$ differs by about 25\% of the pulsation period.  Because we do not consider the energy transfer by radiation, such a phase shift is mainly caused by cooling and shocks. According to criterion 3 of dust formation in Section \ref{sec:dustformation}, we can see from the third panel of Figure \ref{fig:vrprofile} that there are moments when the dust is destroyed at the radius we probe. However, we need to be careful about the temperature variation here because, in reality, \ce{H2} disassociation may absorb a large amount of energy, and the temperature may not rise to the point that sublimates the dust. A full consideration of the realistic EoS is necessary to an accurate model of the dust formation and AGB wind.

\subsection{Morphology of the outflow}\label{sec:morphology}

\begin{figure*}
    \centering
    \includegraphics[width=0.67\columnwidth]{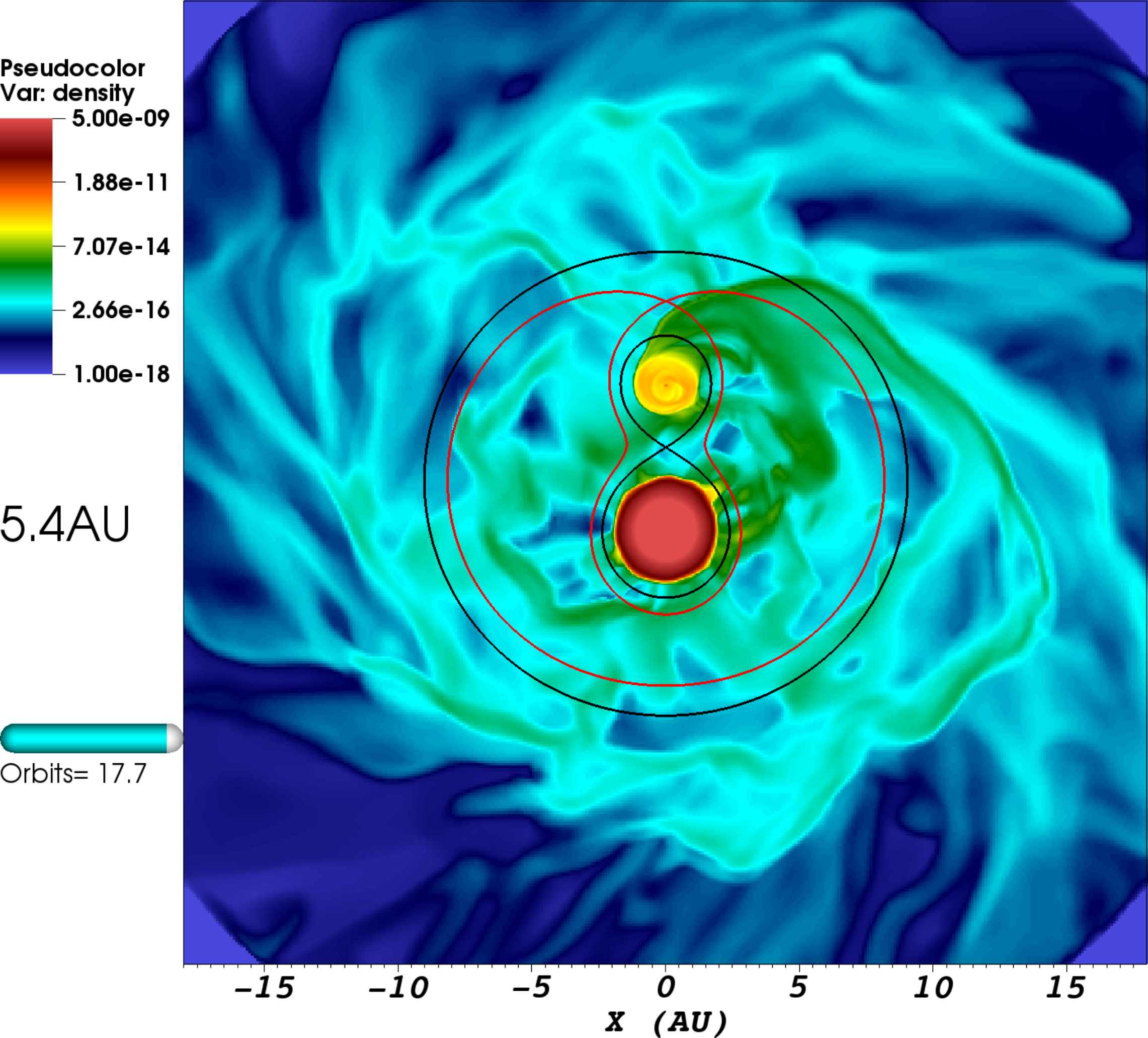}
    \includegraphics[width=0.67\columnwidth]{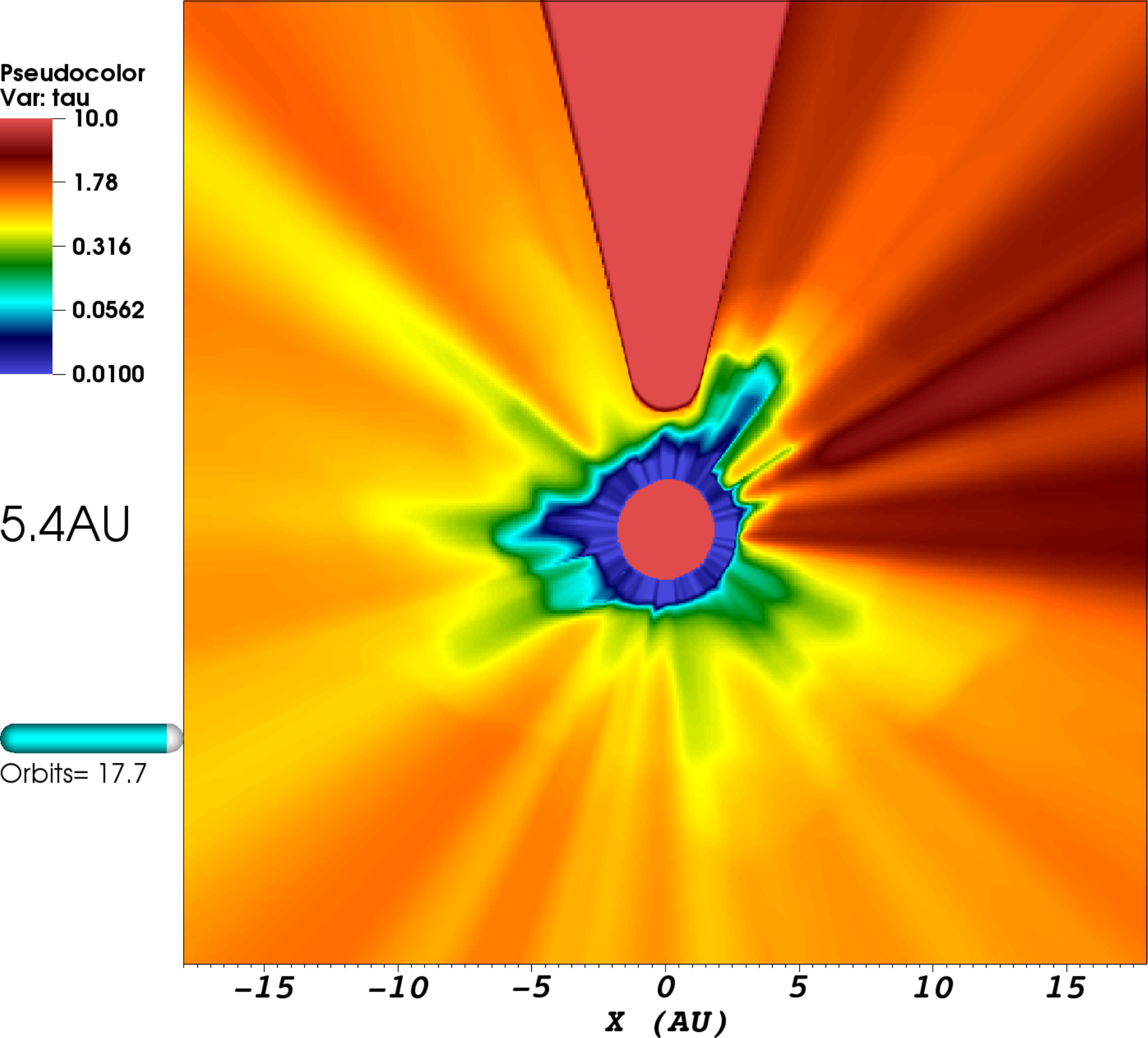}
    \includegraphics[width=0.67\columnwidth]{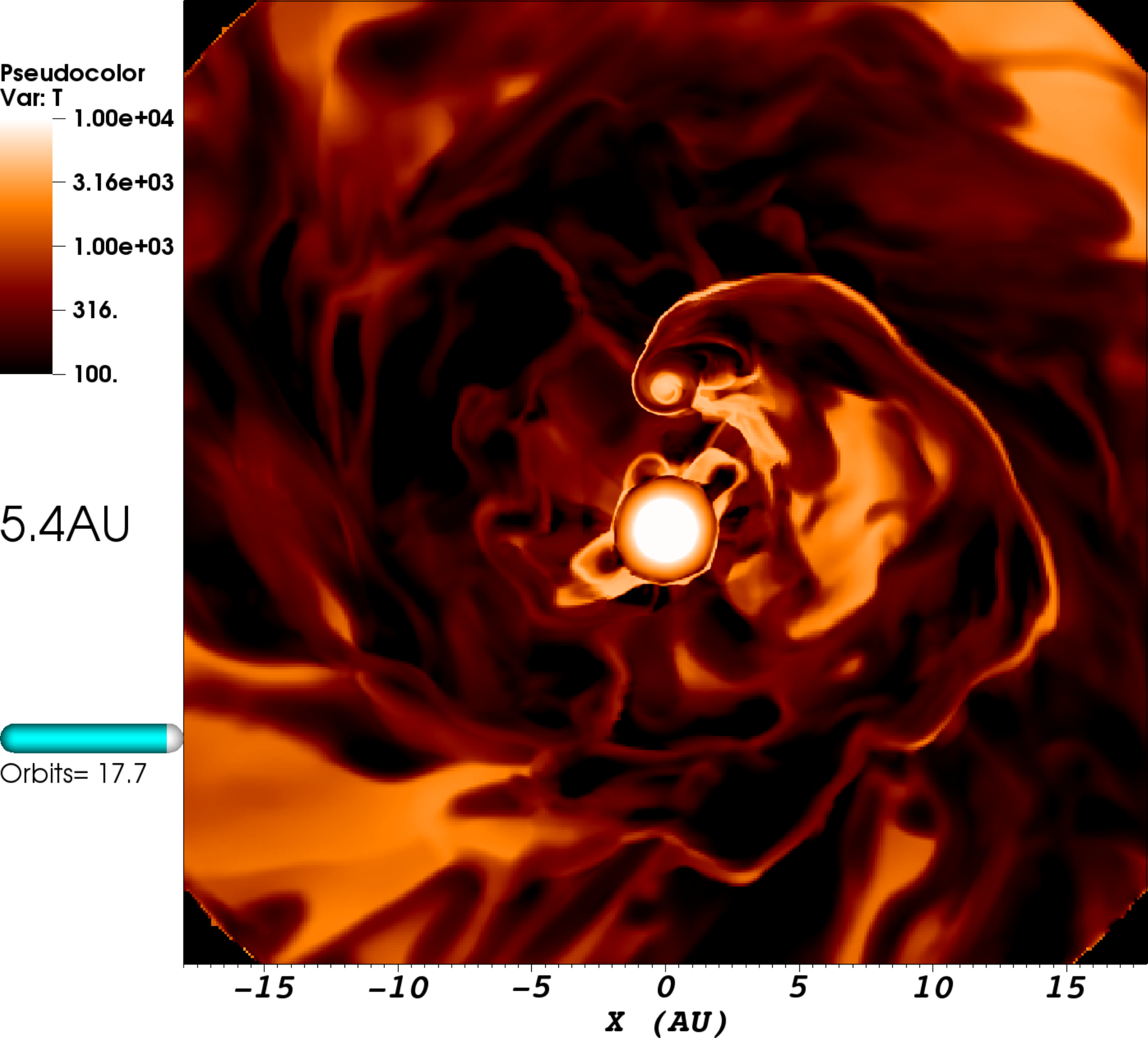}
    \includegraphics[width=0.67\columnwidth]{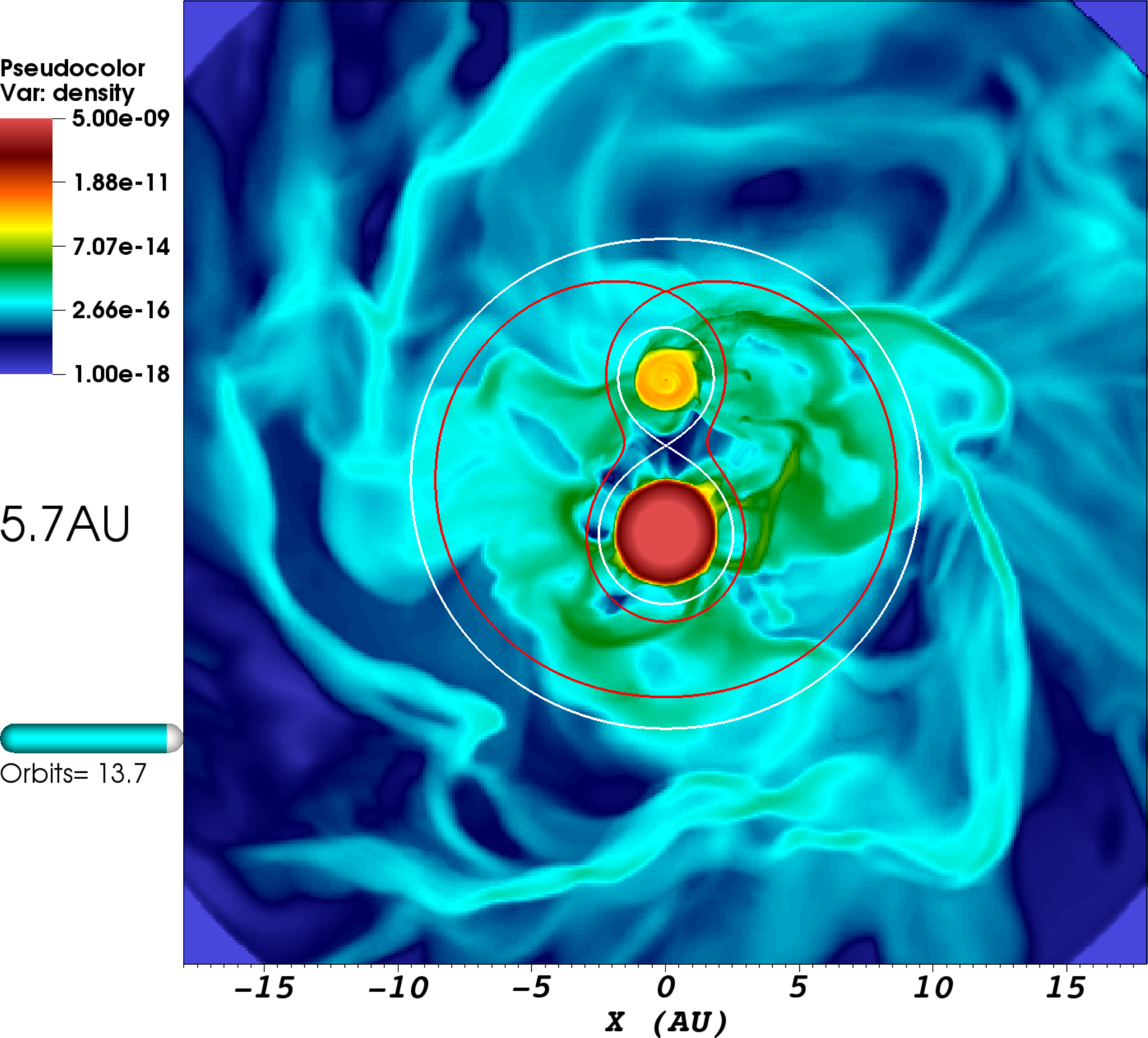}
    \includegraphics[width=0.67\columnwidth]{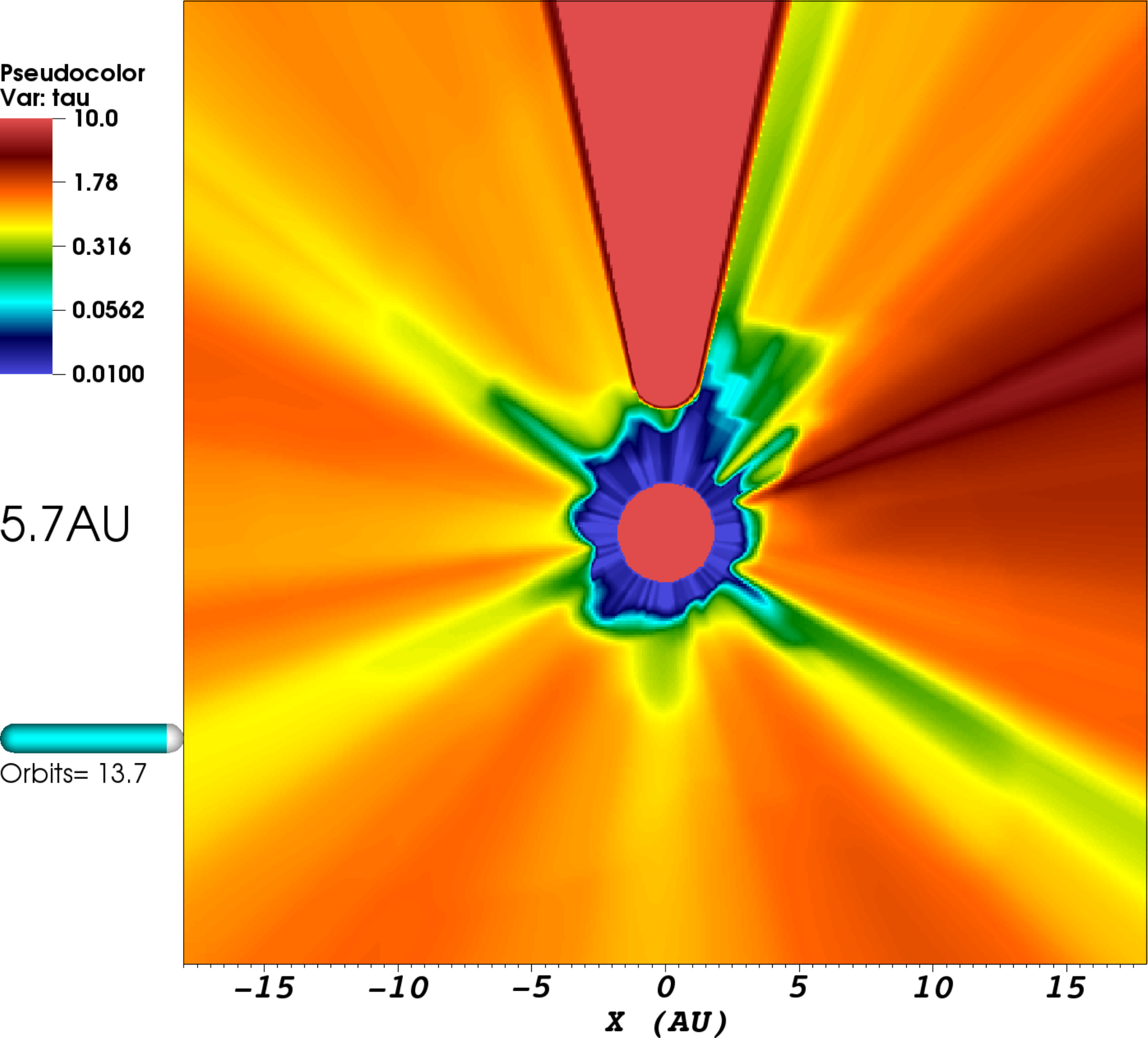}
    \includegraphics[width=0.67\columnwidth]{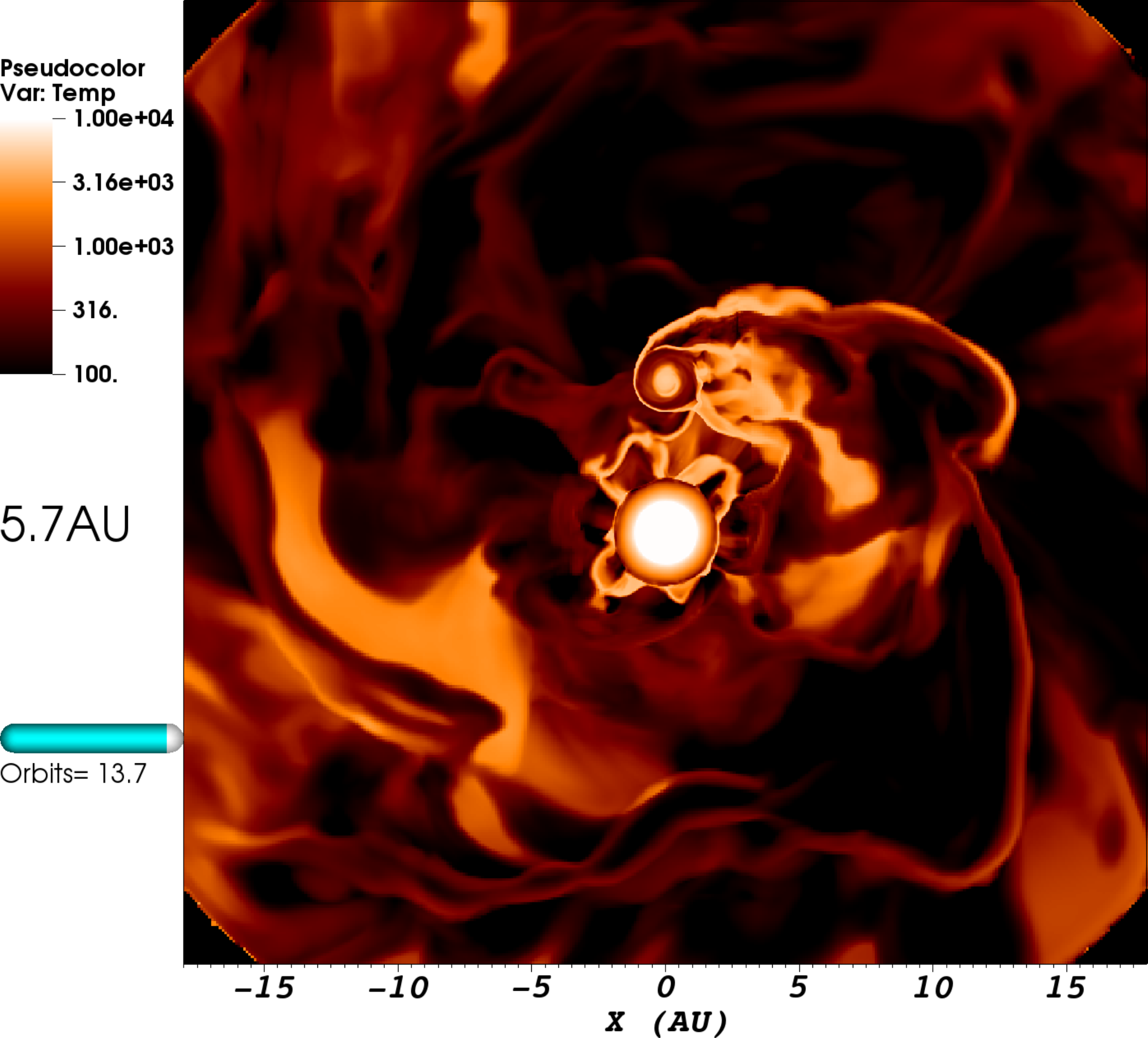}
    \includegraphics[width=0.67\columnwidth]{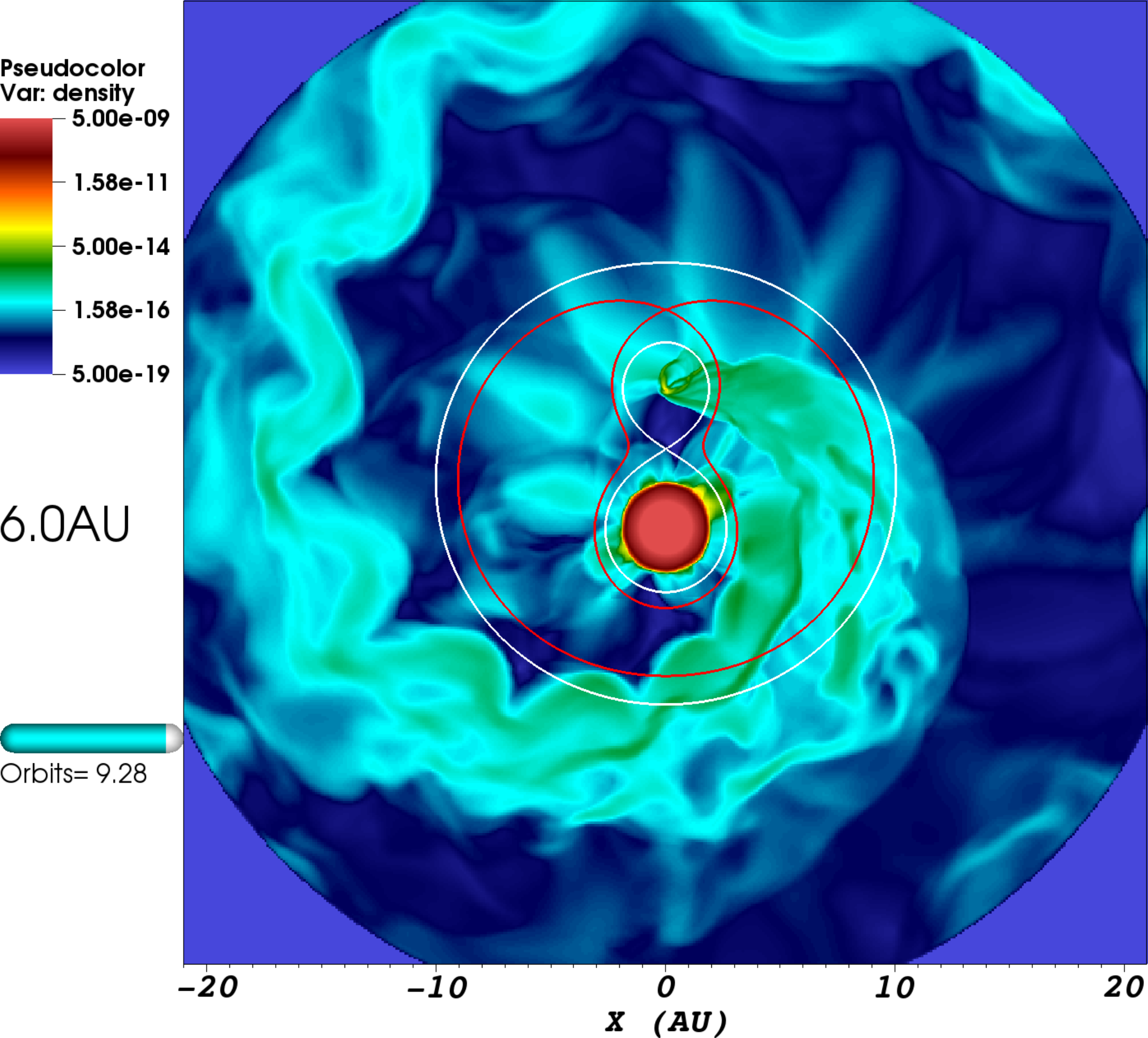}
    \includegraphics[width=0.67\columnwidth]{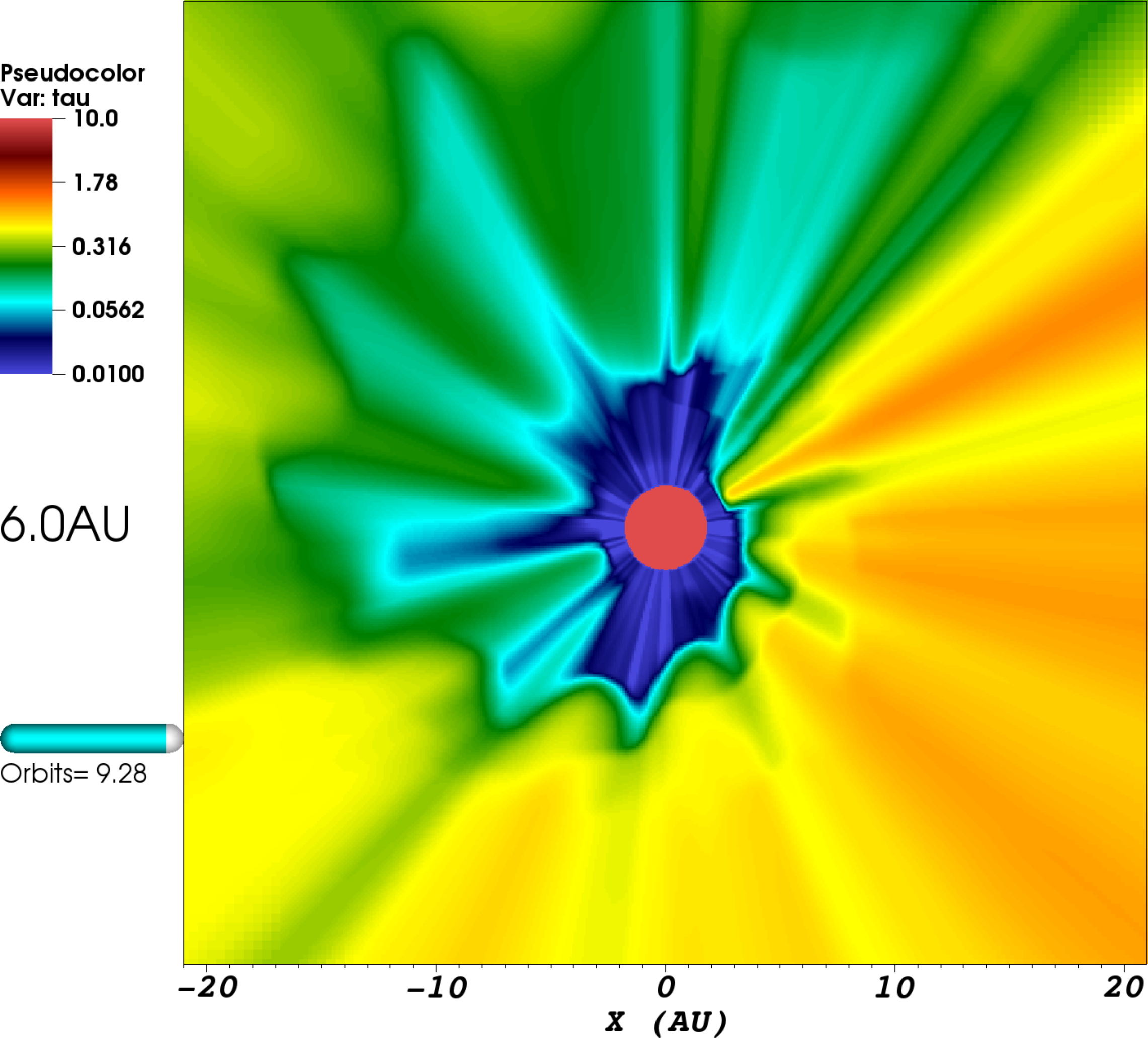}
    \includegraphics[width=0.67\columnwidth]{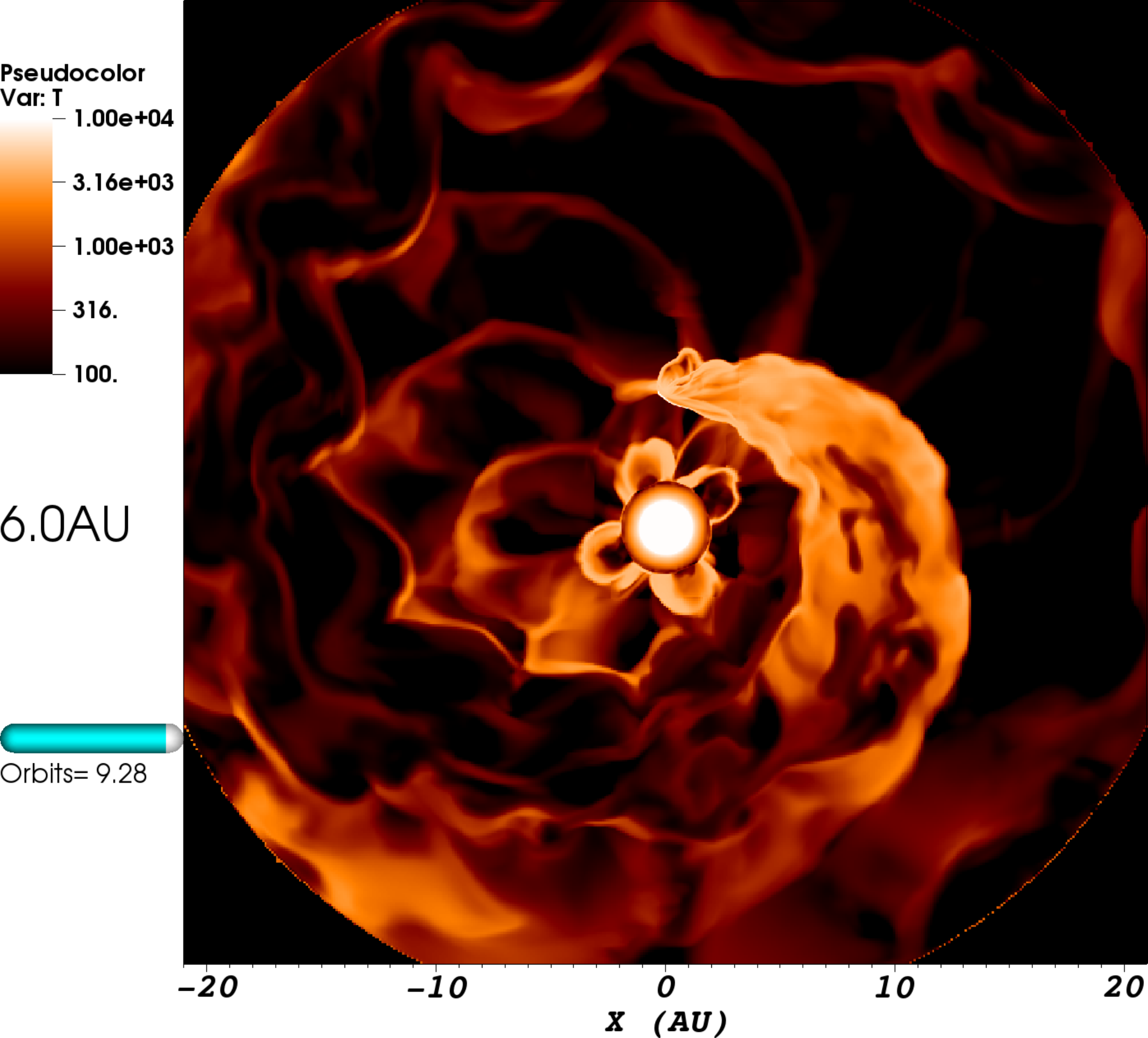}
    \includegraphics[width=0.67\columnwidth]{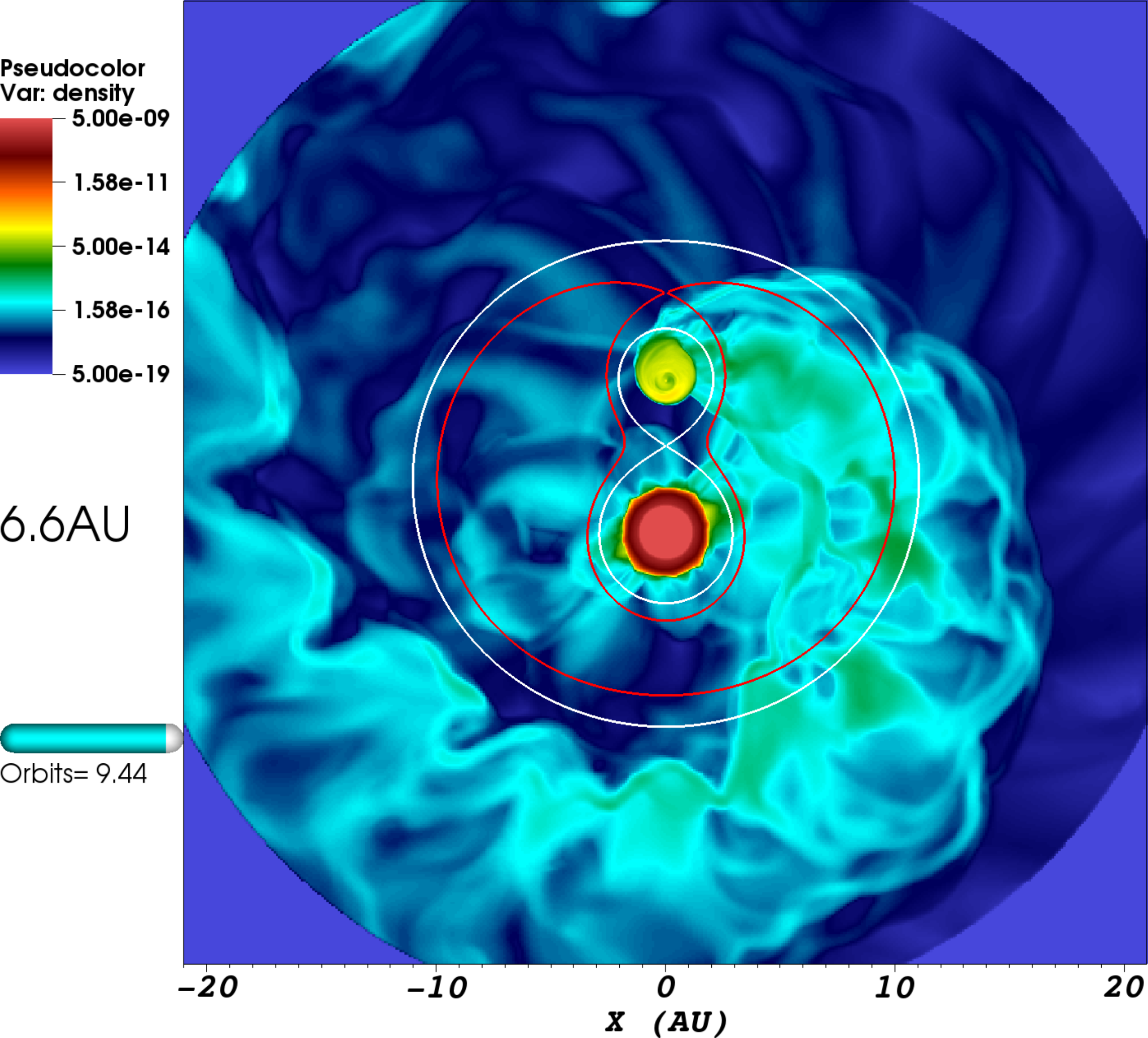}
    \includegraphics[width=0.67\columnwidth]{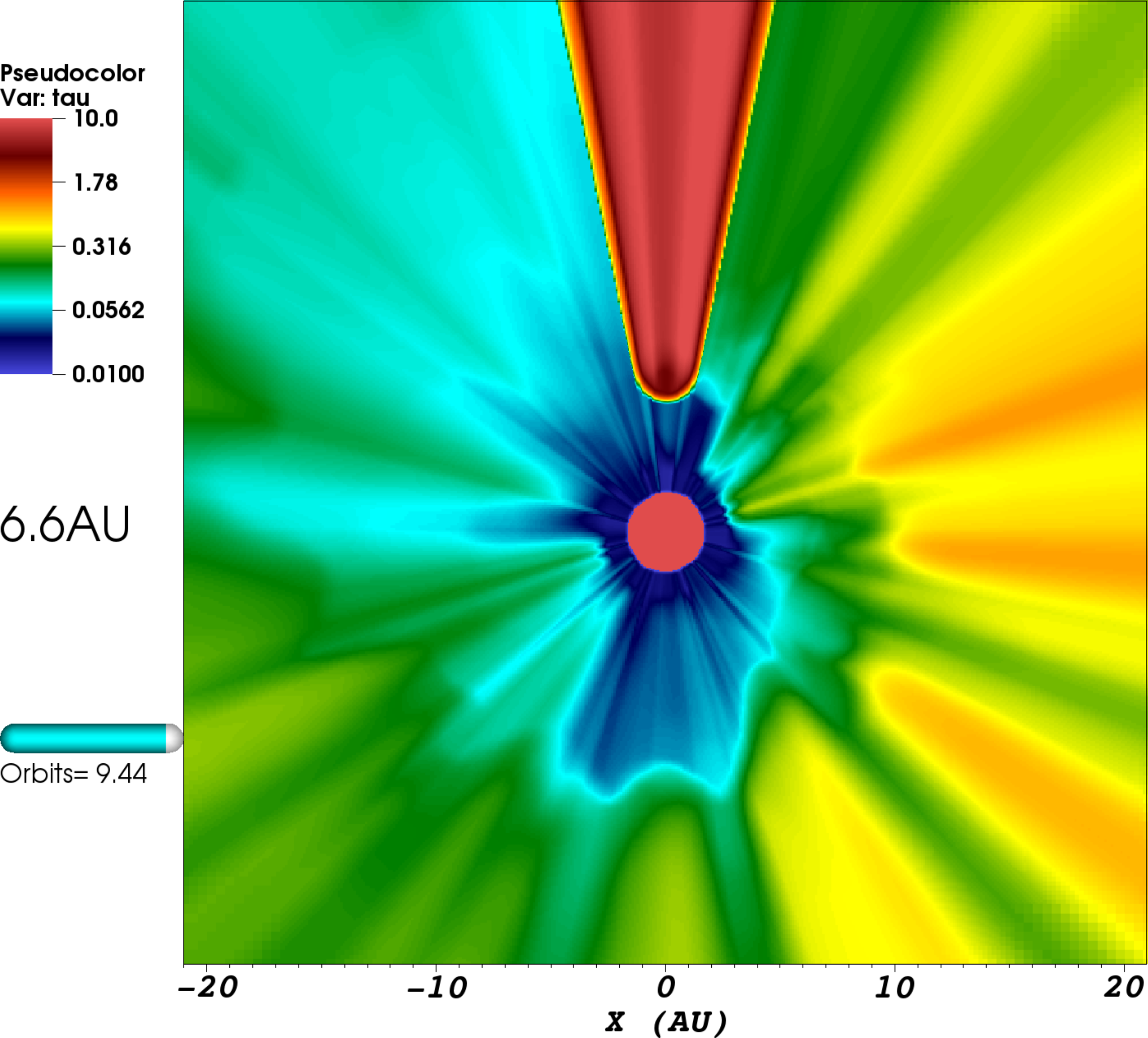}
    \includegraphics[width=0.67\columnwidth]{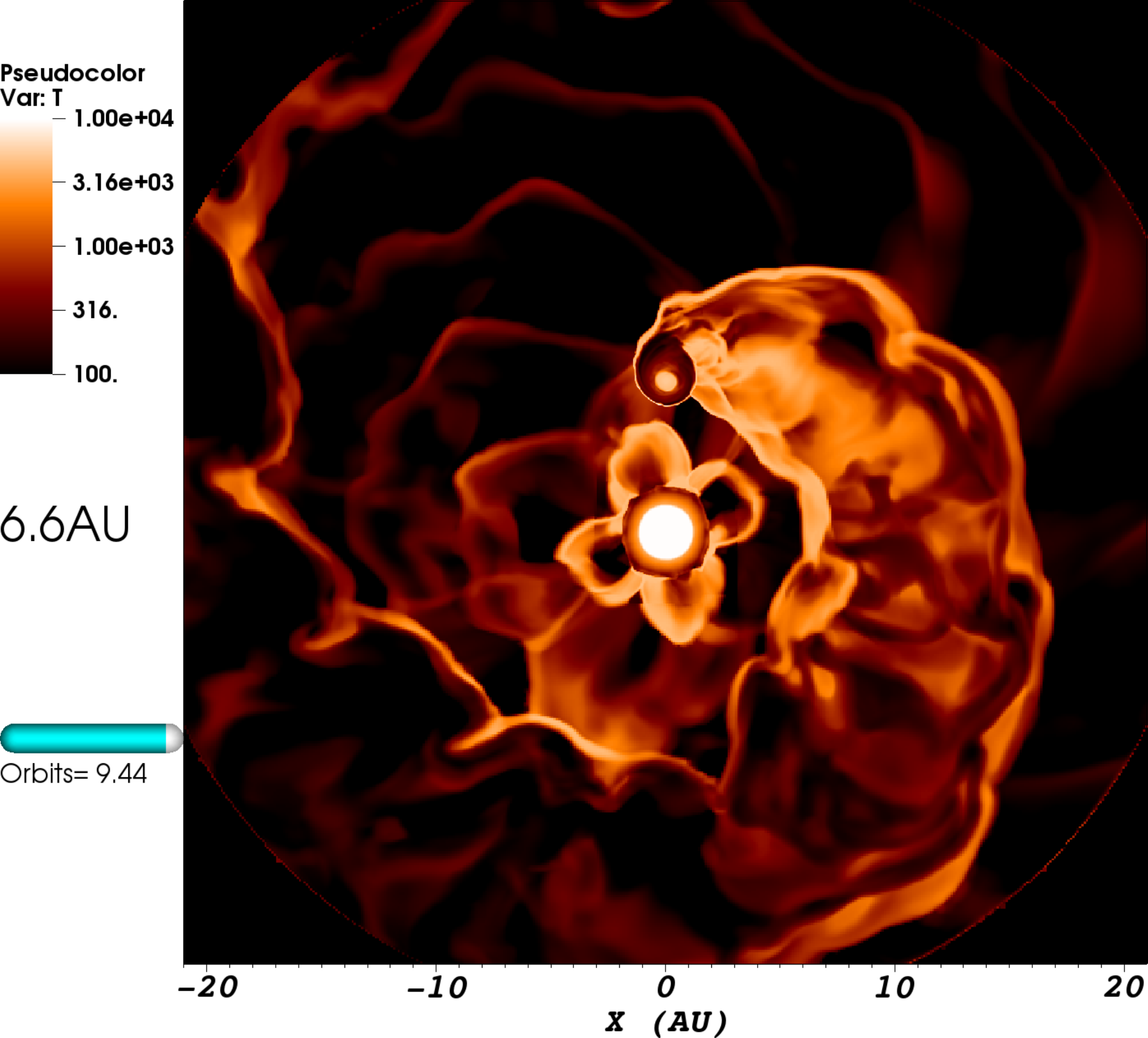}
    \caption{The snapshots for four binaries simulations in $XY$ plane at $Z=0$ for $\rho$ [\gcmc],  $\tau$, and $T$ [K] (from the left to right). From the top to the bottom we show the simulations with $d=5.4,5.7,6.0$ and $6.6$ AU, respectively. The snapshots are provided at 17.7, 13.7, 7.60, and 9.44 orbits, respectively. \new{Due to space limit, we omit the scale of the y-axis. The aspect ratio of all the plots is 1:1.} In the density plots, we also $L_1$ potential with white or black lines and show $L_2$ potential with red lines.}
    \label{fig:outflow}
\end{figure*}

\begin{figure*}
    \centering
    \includegraphics[width=0.9\columnwidth]{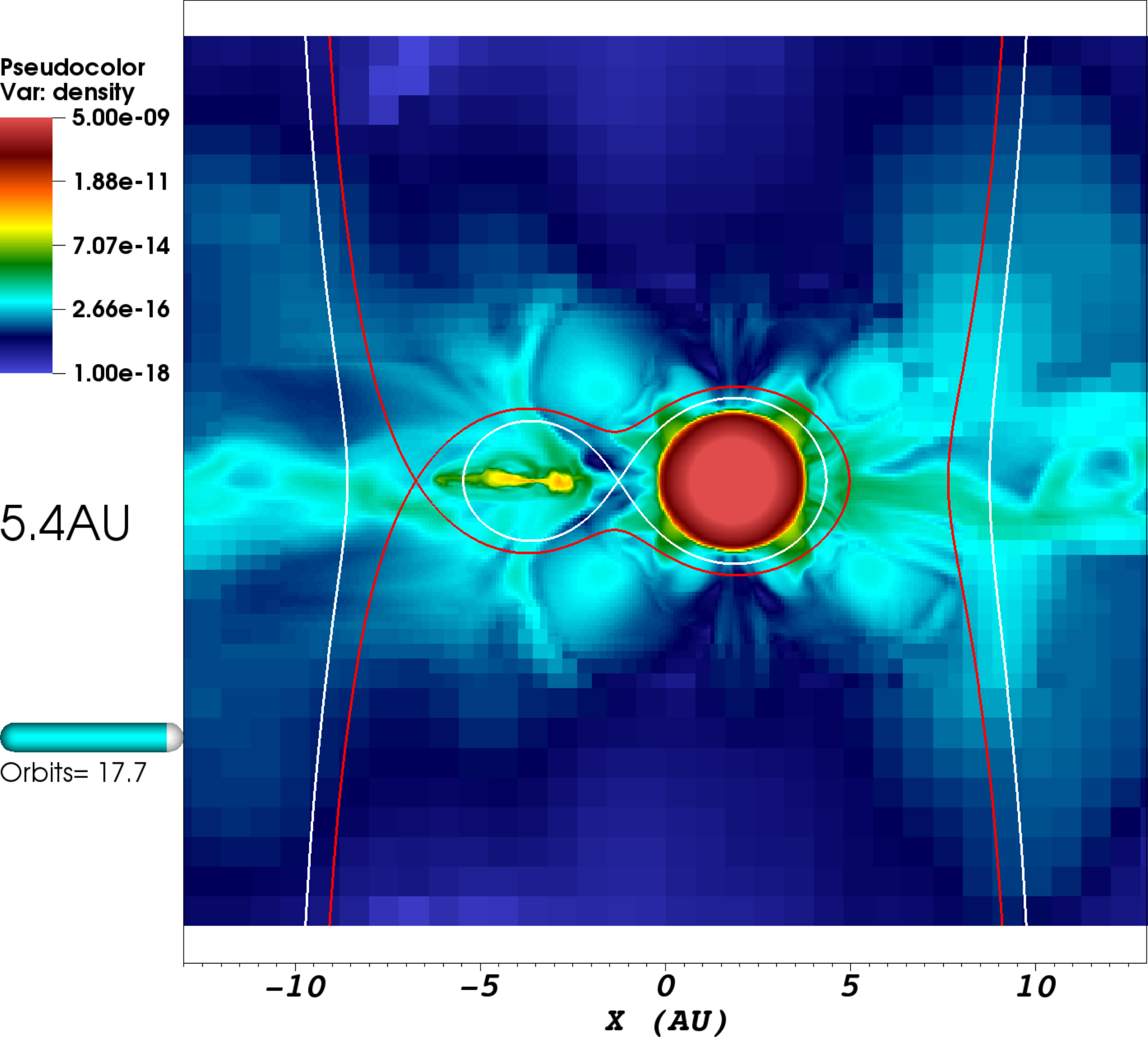}
    \includegraphics[width=0.9\columnwidth]{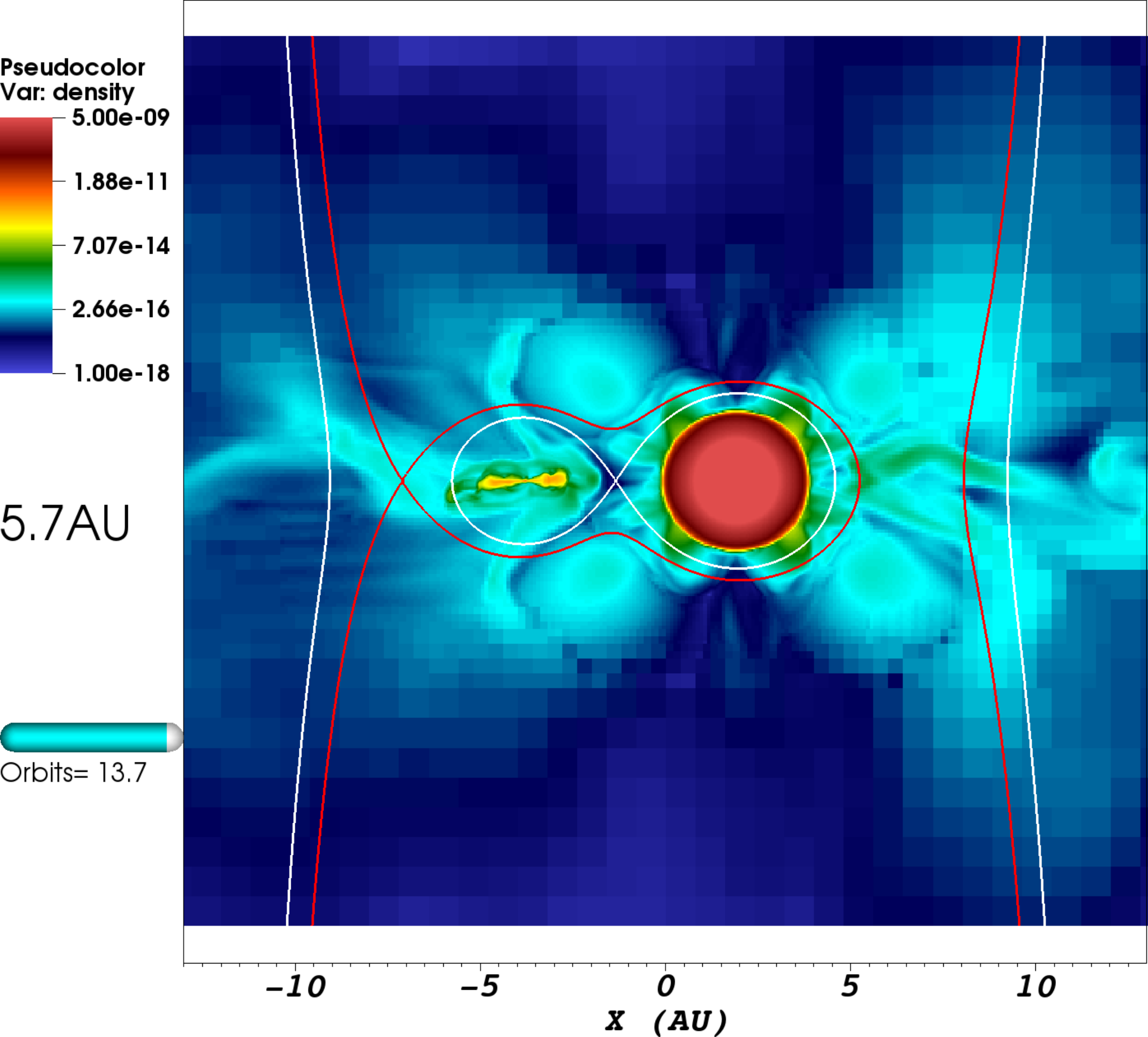}
    \includegraphics[width=0.9\columnwidth]{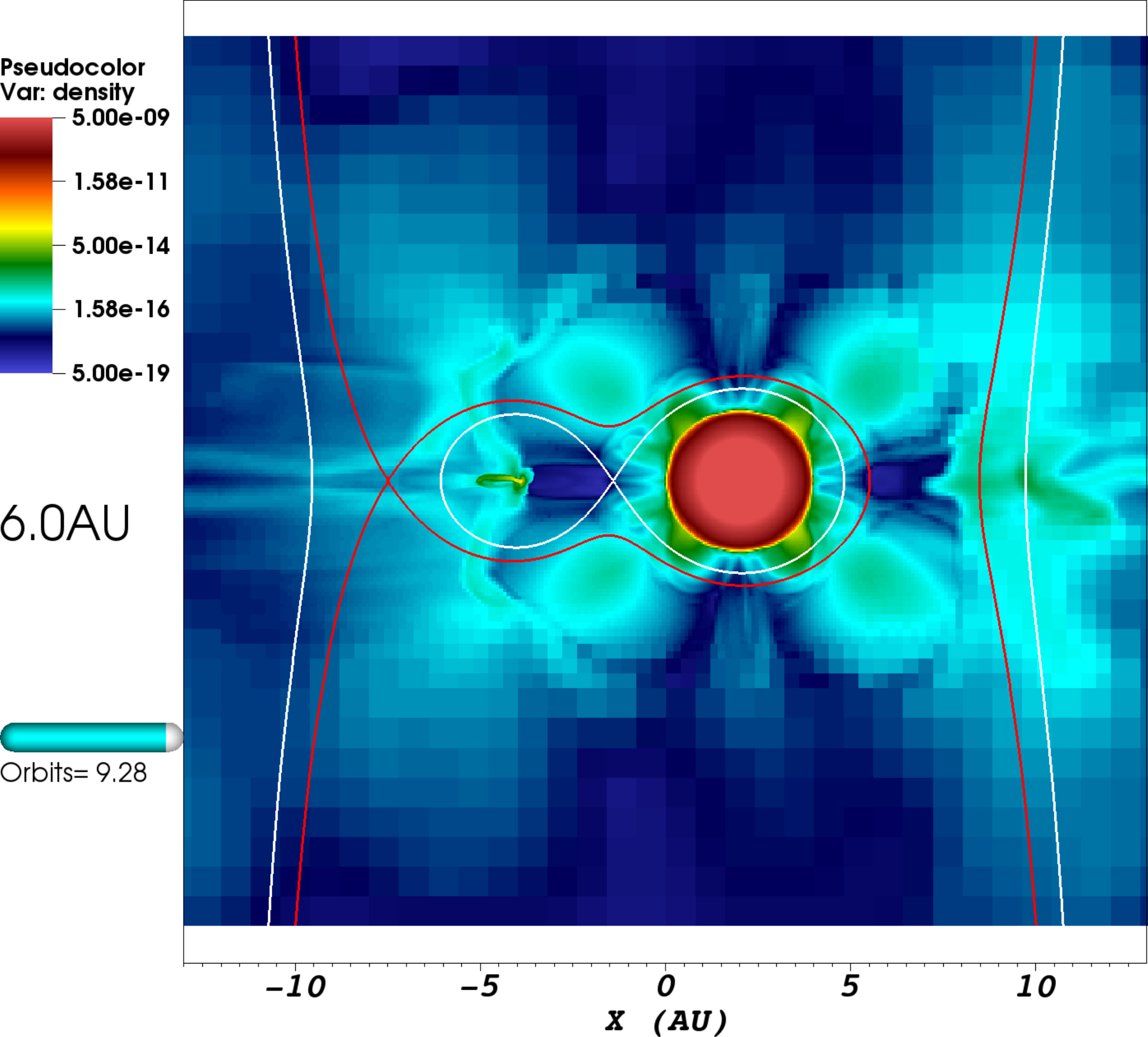}
    \includegraphics[width=0.9\columnwidth]{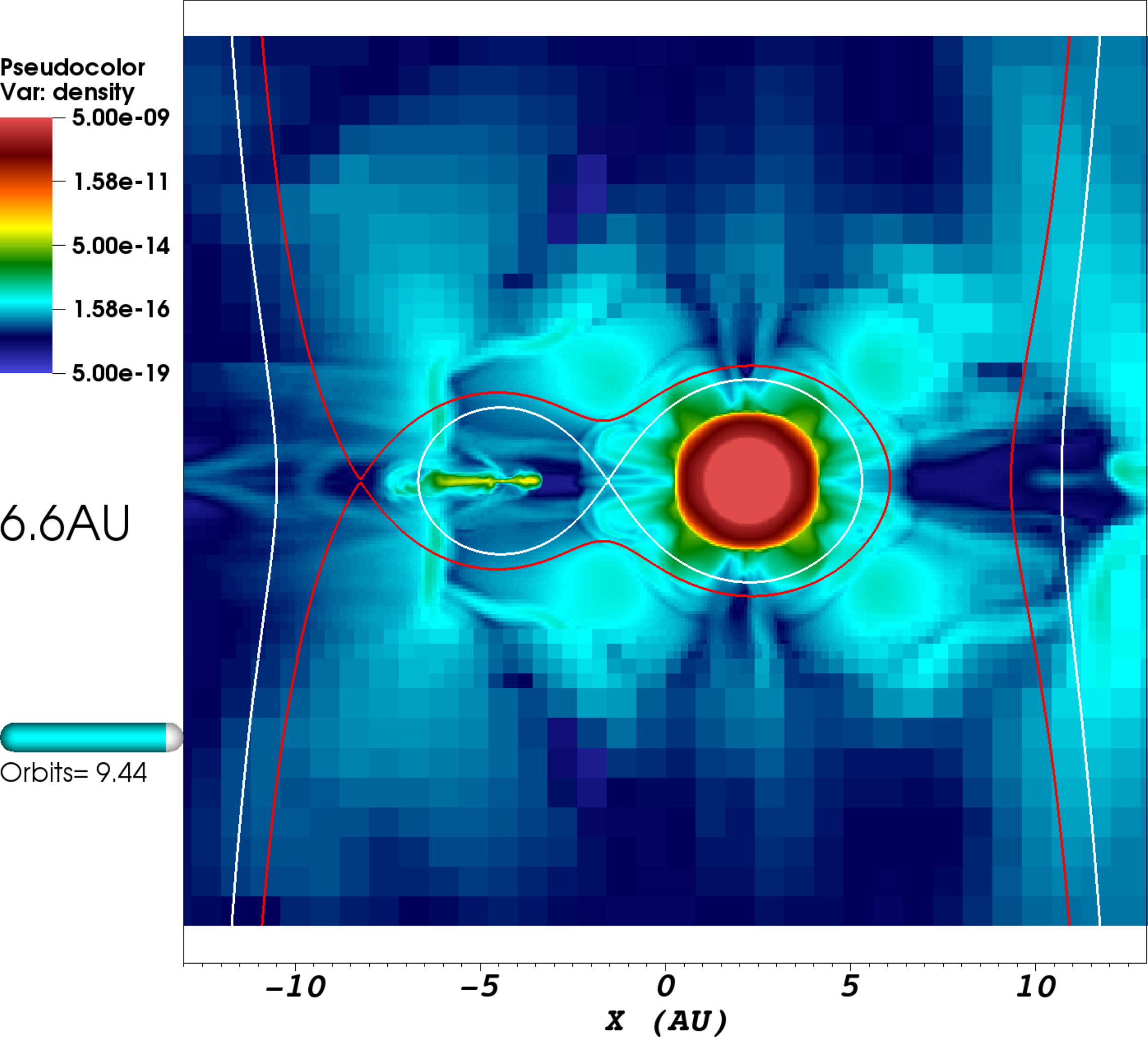}
    \caption{The snapshots for four binaries simulations in $XZ$ plane at $Y=0$  for $\rho$ [\gcmc]. The figure shows the results from simulations with $d=5.4,5.7,6.0$ and $6.6$ AU from the top left panel to the bottom right panel, respectively. Correspondingly, the snapshots are provided at 17.7, 13.7, 7.60, and 9.44 orbits. We also $L_1$ potential with white or black lines and show $L_2$ potential with red lines.}
    \label{fig:side}
\end{figure*}

We present the morphology of the outflows in four binary simulations in Figure \ref{fig:outflow}. The AGB star is in the lower half of each plot. From column 1, we can distinguish two large-scale patterns of outflow structure: the circumbinary disk and the spiral outflow. Specifically, circumbinary disks are found in the two simulations with the smallest separations, while spiral outflows are found in the two widest AGB binaries. We also find these two morphologies in \citetalias{chen2017}.

At a smaller scale, we can distinguish between the appearance or not of the accretion disk around the secondary. Specifically, there is no accretion disk in the simulation with $d=6.0$ AU, while there are thin accretion disks in all the other cases (See Figure \ref{fig:outflow} and \ref{fig:side}). We think the absence of the accretion disk in the $d=6.0$ AU simulation maybe a result of some interplay between the Cartesian mesh and the sub-grid model of the accretion algorithm. In the $d=5.4$ AU, 5.7 AU and 6.6 AU simulations, we can find bow shocks at the edges of the accretion disks (see temperature plots in Figure \ref{fig:outflow}, the shocked material has a higher temperature).

\new{We notice that circumbinary disk forms in the $d=6.0$ AU simulation in \citetalias{chen2017} while it does not form here. Although the binary separation and mass ratio are the same, our current model differs from the model in \citetalias{chen2017} in many aspects. The biggest change that may contribute to the difference in the morphology is the radiation transfer calculation method. In this work, we resolve the radiation force in both polar angle and azimuthal angle, while in \citetalias{chen2017}, only polar angle is resolved. When there is an accretion disk, it is important to model the radiation transfer in different azimuthal angles. For example, if an accretion disk covers $30^\circ$ of the azimuthal angle, and its optical depth is 20. If we average its optical depth over all angles, the optical depth increase will be 1.667, which is big enough to affect the radiation-hydrodynamics in the equatorial plane.}

\new{A former study of AGB binaries shows that the EoS can affect the formation of the accretion disk significantly \citep{theuns1993}. In the adiabatic case (no cooling nor heating), gas with low $\gamma$ EoS can settle onto a thin disk near the accreting object, while the gas with a high $\gamma$ EoS may not be able to form an accretion disk. In this work, as we consider cooling, the effective $\gamma$ is lower than the adiabatic case in the accreting flow, thus a geometrically thin accretion disk is more likely to form.}

We find the optical depth increase drastically when a circumbinary disk present (see middle panel of Figure \ref{fig:outflow}). Each pulsation of the AGB star results in a formation of new clumps in the equator. The set of clumps increases the optical depth irregularly. The accretion disks cast sharp shadows, blocking the radiation from the AGB star.

To qualitatively understand the transition from the spiral outflow structure to the formation of circumbinary disk, we can first consider a spiral outflow that has a finite opacity and a very low density. In this case, the outflow experiences almost the same ratio of $a_{\rm{rad}}/a_{\rm{AGB}}$ where $a_{\rm{AGB}}$ is the gravitational acceleration from the AGB star. The outflow can escape the binary system if $a_{\rm{rad}}$ (or the luminosity of the AGB star) is large enough. If we hold the luminosity of the AGB star constant and increase the density in the spiral outflow, the optical thickness increases, and the value of  $a_{\rm{rad}}$ acting on the outflow material decreases. When the optical thickness increases so fast that $a_{\rm{rad}}$ near the secondary orbit becomes too small to keep the total acceleration positive, the material in the outflow is less likely to escape from the binary system and a circumbinary disk forms. The closer the binary, the more material will be gravitationally focused towards the equatorial plane, and hence a circumbinary disk is more likely to form. In addition, the accretion disk blocks radiation from the primary star. When the material passes behind the disk's shadow, the amount of acceleration it can receive from radiation drops to zero, and the material is more likely to experience a fallback towards binary. In Section \ref{sec:single}, we find that the AGB wind model we adopted has a high wind speed compared to observations and some theoretical models. A high-speed AGB wind is less likely to be captured or deflected by the secondary. Nevertheless, our simulations still show that circumbinary disks form \new{in the two closest binary models}. \new{We anticipate that if a more realistic (i.e., slower) AGB wind is being used in our model, circumbinary disks may form in when the binary separation wider than 5.7 AU.}

In this work, we do not consider the eccentricity pumping and any non-circular motion because the (undetermined) numerical viscosity in our model prevents us from modeling resonances accurately. The interaction between a circumbinary disk and the binary may be one of the sources of the eccentricity \citep{artymowicz1994}.

\subsection{Mass transfer efficiency}\label{sec:beta}

\begin{figure}
    \centering
    \includegraphics[width=\columnwidth]{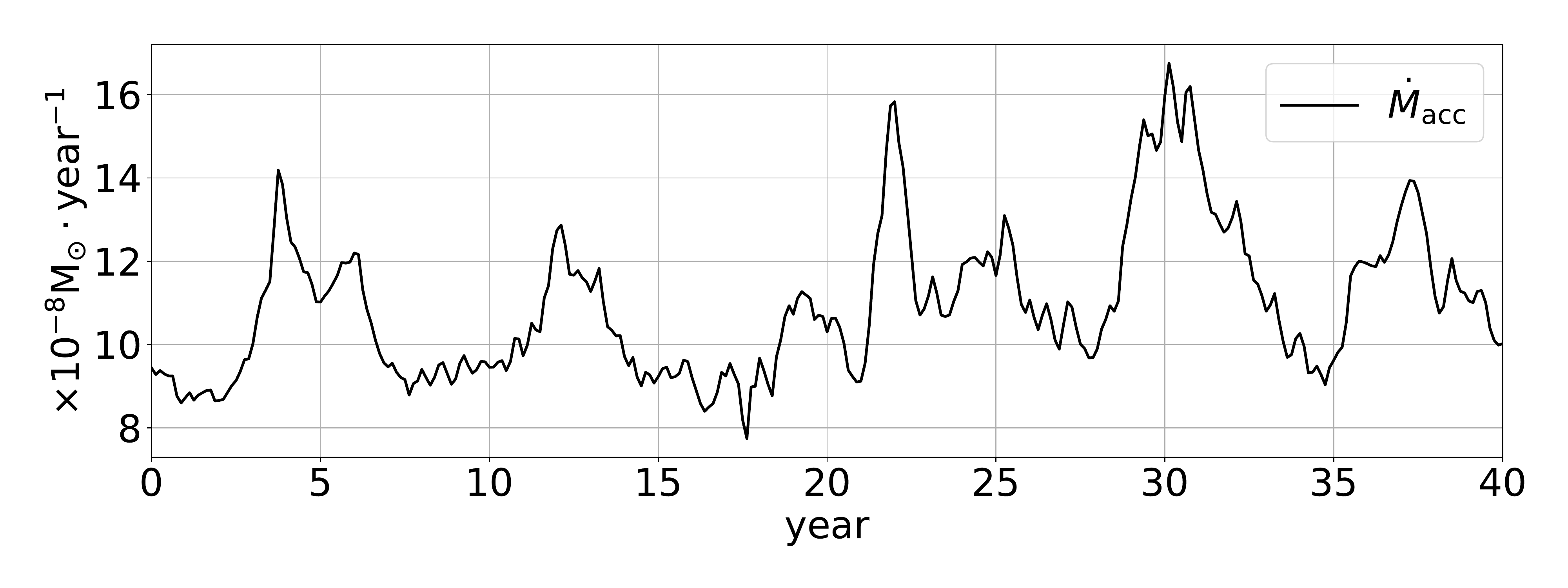}
    \includegraphics[width=\columnwidth]{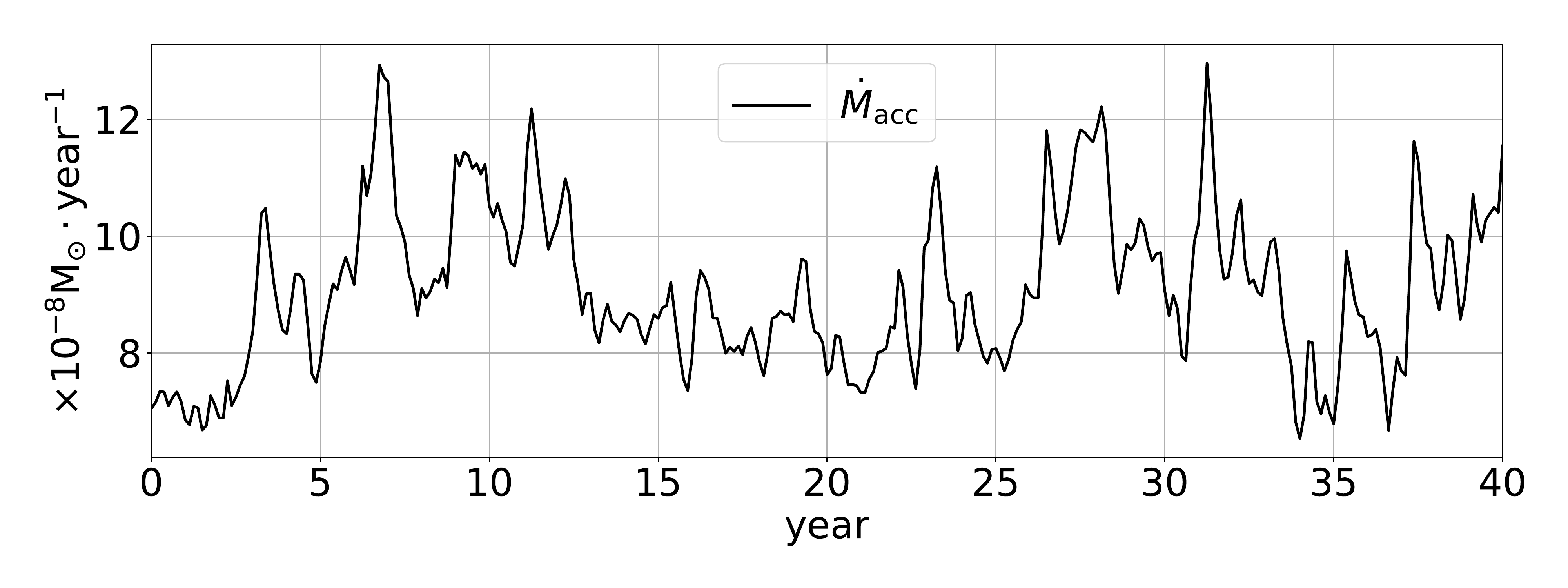}
    \includegraphics[width=\columnwidth]{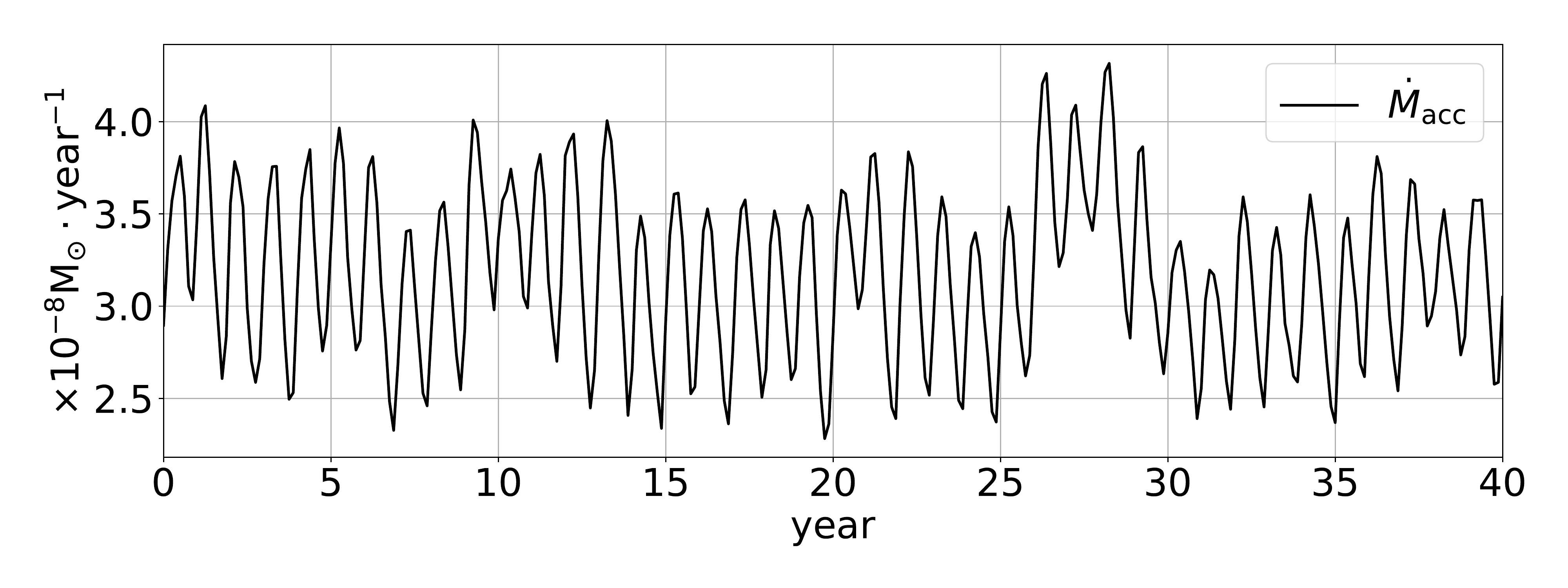}
    \includegraphics[width=\columnwidth]{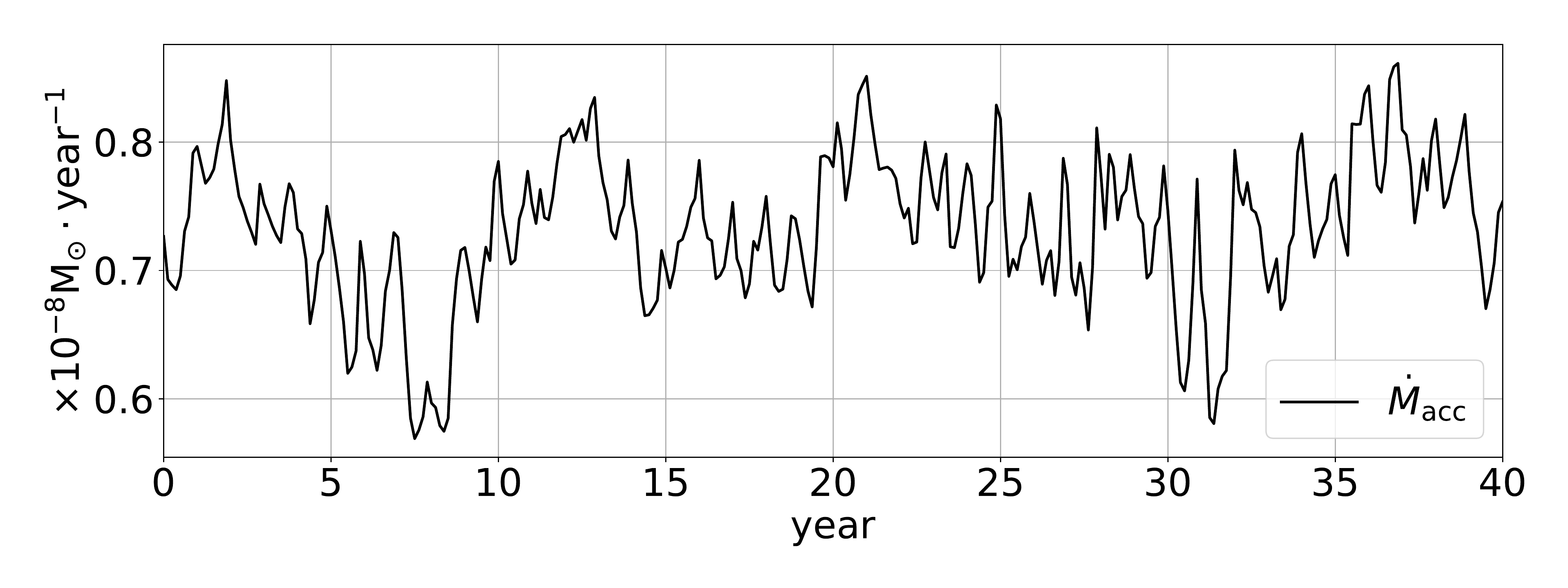}
    \caption{$\dot{M}_{\rm{acc}}$ in the simulations with $d=5.4$ AU, $d=5.7$ AU, $d=6.0$ AU, and $d=6.6$ AU from the top to bottom. The time interval over which the accretion rate is averaged is $0.125$ year.}
    \label{fig:accrate}
\end{figure}

\begin{table*}
    \centering
    \begin{tabular}{ccccccccccc}\hline
    $d$ &   $\dot{M}_{\rm{p}}$  &   $\dot{M}_{\rm{acc}}$ &  $\dot{M}_{\rm{flux}}$  & $\beta$   & $r_{\rm{half}}/r_{\rm{L1}}$  &  $\dot{M}_{\rm{BHL}}$    &   $\beta_{\rm{BHL}}$   &   $\frac{v_{\infty}}{v_{\rm{orbit}}}$   &   morpho  &  mechanism  \\
    $\rm{[AU]}$  &   [\msunyear]   &   [\msunyear]  &   [\msunyear] &   [$\%$]   &   & [\msunyear]   &  [$\%$] \\ \hline
    5.4  &  $3.48\times10^{-7}$  &  $1.08\times10^{-7}$   & $2.40\times10^{-7}$ & 31  &     0.921   &   $1.37\times10^{-8}$    &   3.96    &    1.00    &   CB  &  WRLOF \\
    5.7  &  $3.49\times10^{-7}$  &  $9.09\times10^{-8}$   & $2.58\times10^{-7}$ & 26  &    0.857   &   $1.28\times10^{-8}$    &   3.70    &  1.02  &   CB    &  WRLOF \\
    6.0  &  $4.31\times10^{-7}$  &  $3.18\times10^{-8}$   & $4.00\times10^{-7}$ & 7.4 &    0.814   &   $1.20\times10^{-8}$    &   3.47    &  1.05  &   SO  &  WRLOF    \\
    6.6  &  $4.04\times10^{-7}$  &  $7.37\times10^{-9}$   & $3.97\times10^{-7}$ & 1.8  &  0.740  &   $1.06\times10^{-8}$ &   3.06   &   1.10   &   SO     &  BHL    \\\hline
    \end{tabular}
    \caption{Average mass-loss rate, accretion rate, and mass flux through the sampling shell of the four simulations. $\beta$ is the mass transfer efficiency. The results of the BHL accretion scenario are also calculated. $v_{\infty}/v_{\rm{orbit}}$ is calculated as a reference. Morpho refers to the outflow morphology. CB and SO are short for circumbinary disk and spiral outflow which can be seen in Figure \ref{fig:outflow}.  Mechanism refers to the mass transfer mechanism, in our simulations, it is either the WRLOF or the BHL accretion.}
    \label{tab:result}
\end{table*}

We define the mass transfer efficiency as
\begin{equation}\label{eqn:efficiency}
    \beta=\dot{M}_{\rm{acc}}/\dot{M}_{\rm{p}},
\end{equation}
where $\dot{M}_{\rm{acc}}$ and $\dot{M}_{\rm{p}}$ are the accretion rate of the secondary and the mass-loss rate of the primary. We show the $\dot{M}_{\rm{acc}}$ during the final 40 years of simulations in all models in Figure \ref{fig:accrate}.

We find that the accretion rate varies, and the smaller the distance between the stars, the larger and more irregular is the variation -  $\dot{M}_{\rm{acc}}$ changes by a factor of 2 for  $d=5.4$ AU but only by 30\% for $d=6.6$ AU simulations. We may attribute the irregularity of the accretion rate to the following two reasons.
\begin{enumerate}
    \item The resolution around the secondary may be too low. We use $r_{\rm soft}=0.1$ AU which corresponds to 4 finest cells in our simulation. Low resolution in grid-based hydrodynamics can result in a high and not smooth numerical viscosity. This numerical viscosity in multidimensional hydrodynamic simulations could be larger than the physical viscosity. In an accretion disk, a higher viscosity could lead to a higher accretion rate \citep{frank2002}. Also, the small number of cells of the accretion zone around the secondary may contribute to the irregularity in the accretion \citep{krumholz2004}.
    \item The accretion algorithm is well tested in the BHL accretion scenario \citep{krumholz2004} but is not examined for the case of accreting from an accretion disk. The direct interaction between the accreting particle and the disk around it may be impulsive. \new{In a sub-Keplerian disk}, a sudden removal of some material at the center of the disk may lead to unbalanced forces and create \new{artificial} pressure waves \citep{li2003}. Additionally, we use cooling function to remove the internal energy (and the pressure) inside the disk. The cooled gas is more likely to be accreted according to the accretion algorithm. However, the cooling rate depends on the density and velocity and is highly nonlinear. We anticipate that sometimes the cooling makes a chunk of gas to be less energetic and accreted.
\end{enumerate}

\newr{As we have mentioned in Section \ref{sec:cooling}, the use of the optically thin cooling inside the accretion disks may lead to an overestimate of the cooling rate. The impact of "over cool" depends on how much the accretion disk is pressure-supported. In a disk that is partially supported by the pressure, cooling processes may decrease the pressure gradient of the disk that could lead to a contraction of the disk and a variation in the accretion rate. The amplitude of the variation in the accretion rate depends on the strength of the cooling rate and the pressure gradient of the disk. \citet{saladino2018,saladino2019b} observed variations in the accretion rate in their SPH model with a prescribed cooling rate (their cooling rate may be much less stiff than the one used in this work). We anticipate that the accretion disks in our simulations are partially supported by the pressure, thus, cooling contributes to the accretion rate in our simulation. An over estimation of the cooling rate may eventually lead to an over estimation of the accretion rate and the mass transfer efficiency.}

The accretion rate of $d=6.0$ AU simulation, the one without an accretion disk, is very regular. 
The pulsation period of the AGB star is 1 year, and we find that its accretion rate oscillates at a roughly same period. In Equation \ref{eqn:efficiency}, $\dot{M}_{\rm{p}}$ is the sum of the mass flux $\dot{M_{\rm{flux}}}$ that leaves the binary system and $\dot{M}_{\rm{acc}}$. We find average $\dot{M_{\rm{flux}}}$ by summing the mass flux through a spherical shell with $r_{\rm{shell}}=10$ AU and centered at the origin. We then average $\dot{M_{\rm{flux}}}$ over 40 years, see Table \ref{tab:result}. At the same time we can find what accretion rate can be predicted by the analytic BHL accretion model
\begin{equation}\label{eqn:bhl}
    \dot{M}_{\rm{BHL}}=\frac{G^{2}m_{\rm{s}}^{2}\dot{M}_{\rm{single}}}{v_{\infty}\left(v_{\rm{orbit}}^{2}+v_{\infty}^{2}\right)^{3/2}d^{2}},
\end{equation}
Here $v_{\rm{orbit}}=\Omega_{\rm{b}}d$ is the relative speed between the two orbiting objects. and $v_{\infty}$ is velocity of wind at infinity. \citet{saladino2018} multiplied the RHS of Equation \ref{eqn:bhl} by an efficiency parameter $\alpha_{\rm{BHL}}$ which may vary from 0.8 to 1.5. This parameter is mainly empirical so we do not study which value is more appropriate for which AGB binary model. Readers can multiply $\beta_{\rm{BHL}}$ in Table \ref{tab:result} by a factor to get the corresponding range.

To find what the analytic BHL accretion would predict for our system, we adopt that the AGB star loses mass at the constant rate of $\dot{M}_{\rm{single}}$ and use a constant radial velocity of $v_{\infty}=15.8$\ \kms\ which is the terminal velocity of the AGB wind. Equation \ref{eqn:bhl} is usually the upper limit of the BHL accretion rate because its model ignores all kinds of feedback such as pressure and radiation of the accreting flow. The mass transfer efficiency of the BHL accretion can be found as
\begin{equation}
    \beta_{\rm{BHL}}=\frac{\dot{M}_{\rm{BHL}}}{\dot{M}_{\rm{single}}}=\frac{G^{2}m_{\rm{s}}^{2}}{v_{\infty}\left(v_{\rm{orbit}}^{2}+v_{\infty}^{2}\right)^{3/2}d^{2}}.
\end{equation}
We can see from Table \ref{tab:result} that only the simulation with $d=6.6$ AU  has $\dot{M}_{\rm{acc}}$ and $\beta$ smaller than the BHL model. The simulation with $d=6$ AU has $\dot{M}_{\rm{acc}}$ and $\beta$ larger than the BHL model. At the same time we find that this simulation also exhibits a spiral outflow structure. \new{The high accretion rate suggests that WRLOF is taking place. The spiral structure suggests that the radiation pressure is still large enough to push away the material.}

The two wide AGB binaries exhibit spiral outflow and have a higher mass-loss rate as compared to the two closer ones with circumbinary disks. It is because the circumbinary disk confines some of the gas in the equatorial region and the gas may fall back to the AGB star.

It has been argued that $v_{\infty}/v_{\rm{orbit}}$ is a key parameter of the mass transfer efficiency,  and the orbital evolution of a binary with an AGB star can be predicted by this parameter \citep{saladino2018,saladino2019a}. In our four AGB binary simulations, $v_{\infty}/v_{\rm{orbit}}$ does not change much. However, the mass transfer efficiency changes drastically. The change in mass transfer efficiency seems to be closely related to the formation of the circumbinary disk. The formation of the circumbinary disk is related to the increase of the optical depth in the equatorial region (see Section \ref{sec:morphology}). The formation of a circumbinary disk in close AGB binaries questions the applicability of the simplified mass transfer models.

\subsection{Angular distribution of the outflow}\label{sec:distribution}

\begin{table*}
    \centering
    \begin{tabular}{ccccccc}\hline
    model name      & $f_{1}$&$f_{2}$&$f_{3}$&$f_{4}$&$f_{5}$&$f_{6}$       \\
        &   $[0,30^\circ]$     &   $[30^\circ,60^\circ]$    &   $[60^\circ,90^\circ]$    & $[90^\circ,120^\circ]$    &   $[120^\circ,150^\circ]$    &   $[150^\circ,180^\circ]$ \\\hline
    isotropic   &   $6.70\times10^{-2}$ &   $1.83\times10^{-1}$ &   $2.50\times10^{-1}$ & $2.50\times10^{-1}$   &   $1.83\times10^{-1}$ &   $6.70\times10^{-2}$    \\
    single    &   $5.23\times10^{-2}$ &   $2.19\times10^{-1}$ &   $2.31\times10^{-1}$ & $2.28\times10^{-1}$   &   $2.18\times10^{-1}$ &   $5.26\times10^{-2}$  \\
    $5.4$AU     &   $1.29\times10^{-2}$ &   $1.33\times10^{-1}$ &   $3.62\times10^{-1}$ & $3.54\times10^{-1}$   &   $1.27\times10^{-1}$ &   $1.18\times10^{-2}$  \\
    $5.7$AU     &   $1.43\times10^{-2}$ &   $1.48\times10^{-1}$ &   $3.42\times10^{-1}$ & $3.36\times10^{-1}$   &   $1.47\times10^{-1}$ &   $1.40\times10^{-2}$   \\
    $6.0$AU       &   $1.38\times10^{-2}$ &   $1.18\times10^{-1}$ &   $3.76\times10^{-1}$ & $3.67\times10^{-1}$   &   $1.12\times10^{-1}$ &   $1.33\times10^{-2}$   \\
    $6.6$AU     &   $1.83\times10^{-2}$ &   $1.46\times10^{-1}$ &   $3.36\times10^{-1}$ & $3.44\times10^{-1}$   &   $1.38\times10^{-1}$ &   $1.77\times10^{-2}$   \\\hline
    \end{tabular}
    \caption{The fraction of mass of outflow through binned polar angular range. The second row list the polar angle range of each bin in degree. $0^\circ$ corresponds to the positive z direction. The third row \new{shows the angular distribution if the flux from a sphere} is isotropic. The fourth row shows the result from the single star from Section \ref{sec:single}. The results of the four binary simulations are listed from row 4 to row 8.}
    \label{tab:distribution}
\end{table*}

The secondary deflects the outflow from the primary to the equator, making the shape of the outflow more bipolar \citepalias[see also][]{chen2017}. To quantitatively demonstrate that the secondary focuses the outflow onto the equatorial plane, we split the simulation domain into six bins in the polar angle, where each bin extends for $30^\circ$. We again evaluate the mass flux at $10$ AU from the origin. We find the time average mass of the outflow through each of the bins. We define the fraction of mass of the outflow in each bin as
\begin{equation}
    f_{i}=\dot{M}_{i}/\dot{M}_{\rm{p}},
\end{equation}
where $\dot{M}_{i}$ is the mass of the outflow through the $i$th bin. Values of $f_{i}$ are provided in Table \ref{tab:distribution}. 

Since the secondary is located at the equator, we distribute the accreted mass of the secondary among the two bins in the equatorial region evenly. We find that the outflow of our single star is slightly not isotropic, albeit it is quite symmetric with respect to the polar angle. Its outflow is exceeding a fully isotropic case by about 20\% in the $30-60^\circ$ region. The Cartesian mesh probably causes anisotropy.

From Table \ref{tab:distribution}, one can see that the outflow is strongly concentrated in the equatorial region in all four binary simulations. Let us compare the equatorial bin $f_{3}$ for all four binary simulations. If we compare the two close binaries with the circumbinary disk, the closer one (5.4 AU) has a greater $f_{3}$. Similarly, if we compare the two wide binaries with spiral outflow, the closer one (6.0 AU) has a greater $f_{3}$. If a circumbinary disk is present, some material in the equatorial region may fall back to the AGB star, which can be inferred in the total mass-loss rate $\dot{M}_{\rm{p}}$ in Table \ref{tab:result}.

\subsection{Orbital stability}\label{sec:orbit}

In Section \ref{sec:gravity}, we adopted that the orbit of the binary is circular and does not change. In nature, the binary orbit would change due to the loss of the angular momentum. Here we evaluate the potential rate of the orbital change and check posteriorly if our assumption in Section \ref{sec:gravity} is justified.

The total angular momentum $J$ of the binary system and the specific angular momentum of the binary system $j$ are
\begin{eqnarray}\label{eqn:angularmom}
    J&=&m_{\rm{p}}m_{\rm{s}}\sqrt{\frac{Gd}{m_{\rm{p}}+m_{\rm{s}}}},    \\\label{eqn:specificangularmom}
    j&=&m_{\rm{p}}m_{\rm{s}}\sqrt{\frac{Gd}{(m_{\rm{p}}+m_{\rm{s}})^{3}}}.
\end{eqnarray}
It does not include any stellar spin as in our simulations stars were considered to not rotate. We define $\gamma_{\rm{am}}$ to be the multiple of the specific angular momentum in the outflow in terms of the specific angular momentum of the binary system
\begin{equation}
    j_{\rm{outflow}}=\gamma_{\rm{am}}j.
\end{equation}
It is a measure of the efficiency of angular momentum loss. Making use of $\gamma_{\rm{am}}$, $\beta$, and the mass ratio $q=m_{\rm{p}}/m_{\rm{s}}$, the rate of change of the binary separation can be expressed by
\begin{equation}
    \frac{\dot{d}}{d}=2\frac{\dot{M}_{\rm{p}}}{m_{\rm{p}}}\left(1-\beta q-(1-\beta)(\gamma_{\rm{am}}+\frac{1}{2})\frac{q}{1+q}\right).
\end{equation}
We have calculated $\dot{M}_{\rm{p}}$ and $\beta$ of each binary simulation in Section \ref{sec:beta}. The only unknown is $\gamma_{\rm{am}}$. We get $\gamma_{\rm{am}}$ from the binary simulations by evaluating the average $\gamma_{\rm{am}}$ through a spherical shell whose radius is $1.3d$ and center is at the center of mass of the binary. By setting the sampling radius at $1.3d$, we probably underestimate the angular momentum loss from the binary because the escaping gas can still gain angular momentum as it goes beyond $1.3d$ \citep{lin1977,saladino2018}. However, the angular momentum conservation becomes worse in our code as the radius goes large \citep{chen2018}. We prefer not to incur too much uncertainty, so we set the sampling shell small. The specific angular momentum of each star are,
\begin{equation}
    j_{\rm p,s}=m_{\rm s,p}^{2}\sqrt{\frac{Gd}{(m_{\rm{p}}+m_{\rm{s}})^{3}}}
\end{equation}
Therefore, by taking Equation \ref{eqn:specificangularmom} into consideration, we can find two limiting cases for $\gamma_{\rm am}$. When $q=2$ and $\gamma_{\rm{am}}=0.5$, the gas emitted from the AGB star does not gain any angular momentum, when $\gamma_{\rm{am}}\approx2$, the escaping gas has a specific angular momentum that similar to the secondary. In Table \ref{tab:orbit},  we list values of $\gamma_{\rm am}$ for each of our simulations.  The obtained values are in a between of the two extreme cases. As the distance increases, $\gamma_{\rm am}$ decreases towards the specific angular momentum of AGB star.

At the risk of introducing too many variables (readers will appreciate this variable later), we convert the $\gamma_{\rm am}$ derived from our simulations to a new quantity defined by
\begin{equation}
    \eta=\frac{q\gamma_{am}}{(1+q)^2}=\frac{Jm_{\rm p}m_{\rm s}}{(m_{\rm p}+m_{\rm s})^2}.
\end{equation}
The Equation 7 of \citet{saladino2019a} is an empirical formula that uses $q$ and $v_{\infty}/v_{\rm orbit}$ to predict the value of $\eta_{\rm em}$. Table \ref{tab:orbit} lists $\gamma_{\rm{am}}$, $\eta$, and $\eta_{\rm em}$ of each of our binary simulation and the potential rate of change of the orbit. It turns out that $\eta$ derived from each of our simulation is indeed close to $\eta_{\rm em}$. The match suggests that even though $v_{\infty}/v_{\rm orbit}$ may not be effective in characterizing the mass transfer efficiency when a circumbinary disk is about to form (see Section \ref{sec:beta}), $v_{\infty}/v_{\rm orbit}$ could be useful in predicting the angular momentum loss from the AGB binary.
\begin{table}[]
    \centering
    \begin{tabular}{ccccccc}\hline
    $d$ &  $\gamma_{\rm{am}}$   &  $\delta\gamma_{\rm{am}}$ &   $\eta$ &   $\eta_{\rm em}$  &  $\dot{d}/d$ & $\Delta d$   \\
    $\rm{[AU]}$    &    & [\%]  &   &   &   [$\times10^{-7}/$yr] & [AU] \\\hline
    $5.4$ &   0.955 & 91.0  &   0.212   &   0.199   &   $-1.98$    &   $-1.07$    \\
    $5.7$ &   0.946 & 89.2  &   0.210   &   0.196   &   $-1.60$    &   $-0.911$    \\
    $6.0$ &   0.879 & 75.8  &   0.195   &   0.190   &   $0.0103$    &   $0.0062$    \\
    $6.6$ &   0.815 & 63.0  &   0.181   &   0.181   &   $0.816$    &   $0.538$    \\\hline
    \end{tabular}
    \caption{The measured $\gamma_{\rm{am}}$ of the four binary models and relative rate of the change of the binary separation. $\delta\gamma_{\rm{am}}=((\gamma_{\rm{am}}-0.5)/0.5)\times100\%$ is the percentile difference of the specific angular momentum of the outflow compared to the specific angular momentum of the AGB star. The seventh column show the potential change in $10^{6}$ years if the binary separation change rate is kept constant.}
    \label{tab:orbit}
\end{table}

From Table \ref{tab:orbit}, we can infer that the binary separation in our four binary simulations should not change more than $5\times10^{-4}$ AU. Therefore, our assumption in Section \ref{sec:gravity} is reasonable. On the other hand, if the binary separation change rate is kept constant for $10^{6}$ years, \new{which corresponds to the average lifetime of an AGB star}, the decrease in binary separation of the two close binary is non-negligible. Considering that we have ignored the spin-orbit coupling of the binary and we set a small sampling shell to get $\gamma_{\rm{am}}$, we are underestimating the orbital angular momentum loss. The actual orbital shrinkage should be higher. As the binary separation decreases, the WRLOF may become more similar to the RLOF.

\section{Conclusions and Discussions}\label{sec:conclusion}

In this paper, we carry out 3D radiation-hydrodynamic simulations of AGB binaries. We use ray-tracing radiation transfer to resolve the optical depth in the azimuthal and polar directions. We take into account that dust can be destroyed by shocks, in a high temperature close to an LTE environment, and if radiation from the AGB star can sublimate the small dust grains (see Section \ref{sec:radforce} for detail). We calculate the cooling strength on \ce{H2}, \ce{H}, and \ce{H+}; their number densities are found by solving the Saha equations \citep{chen2019}. We model the AGB star as an inner boundary condition that has sinusoidally varying radial velocity \citep{bowen1988}. We use {\tt MESA} to obtain the temperature and density that determine the boundary conditions of our piston model.

The single star model presented in Section \ref{sec:single} has an average mass-loss rate of $3.45\times10^{-7}$ \msunyear\ and a terminal wind speed of $15.8$ \kms. We find that the mass-loss rate is reasonable for an M-type AGB star with an effective temperature of $2874$ K and a luminosity of $4384$ \lsun \citep{hofner2018}. Observations show that the terminal wind speed should be 7-13 \kms\ \citep{hofner2018}. We argued that this difference would not adversely affect our conclusion. On the contrary, lower terminal wind speed may be captured or deflected by the secondary more easily. In the case of low wind speed, a circumbinary disk may form when the binary separation is greater than 6.0 AU (see Section \ref{sec:morphology} for detail).

We find that the accretion rate increases when the binary separation decreases (see Table \ref{tab:result}). The mass transfer efficiency may increase dramatically to $31\%$ when a circumbinary disk forms. Such a mass transfer efficiency can be up to 8 times of what the canonical BHL accretion rate predicts. The presence of a circumbinary disk imposes severe challenges to the simplified mass transfer mechanisms. It enlarges the domain of dependence of the secondary to the whole equatorial region. The circumbinary disk may also exert a torque on the binary and change the eccentricity of the binary system \citep{artymowicz1994,dermine2013,moody2019}. The eccentricity problem has been heavily studied by the proto-planetary disk community, and similar conditions may apply here \citep{ragusa2018}. The formation of a circumbinary disk is a result of the focused fluids in the equatorial plane by the secondary (Section \ref{sec:distribution}). The focused fluids increase the optical depth in the equatorial plane; thus, the radiation pressure on the fluids decreases. The balance between the radiation pressure and the gravity will be tilted toward the gravity when the optical depth becomes larger than a critical value, and some of the gas may fall back. \new{We point out that the non-local momentum transfer by radiation is crucial in the formation of the circumbinary disks.} \newrr{It would be difficult to observe the formation of the circumbinary disks without the non-local momentum transfer even when WRLOF takes place \citep{mohamed2012}.}

The accretion rates in the two closest AGB binaries are irregular. We outlined two possible reasons for the irregularity in Section \ref{sec:beta}. It may be worthwhile to increase the resolution and allow a subsonic atmosphere (or envelope) to build up around the secondary in future researches. The atmosphere can provide a subsonic region that may smooth out the supersonic flow that falls onto the secondary. The particle can accrete mass from the atmosphere when the atmosphere becomes Jeans unstable \citep{jean1902,federath2010}. It is also natural to model proper feedback, i.e., radiative cooling, from the secondary when an atmosphere presents. In our current model, we have not taken radiative feedback from the accretor and the radiation transfer in the accretion disk into consideration. In a realistic circumstance, the accretion disk will become optically thick when the density of the disk becomes high enough. The transition from the optically thin to the optically thick state may incur different accretion modes. A better understanding of the accretion rate may be achieved by modeling the transition correctly.

The two closest binaries have dusty circumbinary disks, which may resemble the situation AR Puppis \citep{ertel2019}. The two widest binaries have spiral structure outflows. Such structure resembles many post-AGB binaries with wide binary separations or planetary nebulae \citep{edgar2008,maercker2012,ramstedt2017,kim2019}.

In Section \ref{sec:orbit}, we discussed the orbital dynamics of the AGB binaries in our simulations. We confirmed that it is reasonable to ignore the orbital evolution in our simulations. We also find that the two closest AGB binaries may experience an orbital shrinkage, where the binary with the initial separation of $5.4$ AU has the fastest orbital separation decreasing rate. Since we did not consider the tidal spin-orbit coupling and excluded the contribution of angular momentum transfer beyond $1.3$ binary separation from the center of mass in the calculation of $\gamma_{\rm{am}}$, we are underestimating the orbital angular momentum loss in close binaries. The decrease of the binary separation in the two close binaries should be faster than what we estimated. The binary with $5.4$ AU initial separation may become RLOF as it evolves.

When there is no circumbinary disk, the longest timescale in AGB binaries is typically the binary's orbit. When a circumbinary disk presents, the overall conditions (density, chemistry, and timescale) become similar to a proto-planetary disk. The temperature in the circumbinary disk may change a lot because of the shocks and radiation from the AGB star. The material in the equatorial region may stay close to the binary for several or more orbits. The dense environment ($10^{-15}-10^{-14}$\ \gcmc\ in the middle plane near the binary orbit) in the circumbinary disks can also foster dust growth. Large dust grains may process photons with a wavelength longer than the K band. If multiple scattering by the dust becomes essential, a more sophisticated non-gray radiative transfer model is in need. In our simulations, we find that the optical depth as calculated along the line of sight from the AGB star is typically less than 10 (except for the accretion disk). Monte Carlo radiative transfer would be an efficient way to model the radiation-hydrodynamics in the circumbinary disk.

In summary, the mass transfer efficiency $\beta$ and angular momentum loss efficiency $\gamma_{\rm{am}}$ are two important quantities but with large uncertainty in AGB binaries. They hugely affect the binary evolution. In this work, we derived $\beta$ and $\gamma_{\rm{am}}$ from 3D radiation-hydrodynamic simulations. We discover a huge discrepancy (up to 8 times) in $\beta$ by comparing our simulations with \new{the canonical BHL accretion mechanism}. The discrepancy is closely related to the presence of the circumbinary disks. It is essential to carry out the non-local radiation transfer calculation to model the formation and dynamics of a circumbinary disk.

\software{{\tt AstroBEAR} \citep{carroll2013}, {\tt matplotlib} \citep{hunter2007}, {\tt MESA v10398} \citep{paxton2011,paxton2013,paxton2015,paxton2018}, {\tt Visit 3.0.1} \citep{HPV:VisIt}.}

\acknowledgments
We thank the anonymous referee whose suggestions have greatly improved the quality of this work. We thank Craig Heinke, Rodrigo Fernandez, Xuening Bai, Falk Herwig, Pavel Denisenkov, Nami Mowlavi, Ue-Li Pen, and Dong Lai for inspirational discussions. ZC is grateful to the CITA National Postdoctoral Fellowship and the Tsung-dao Lee visiting scholarship.
N.I. acknowledges support from CRC program and funding from NSERC Discovery.
This research was enabled by the use of computing resources provided by Compute/Calcul Canada, and was supported in part by the National Science Foundation under Grant No. NSF PHY-1748958.

\bibliography{references}

\appendix
\section{Optical depth}\label{sec:tau}
\begin{figure}
    \centering
    \includegraphics[width=\columnwidth]{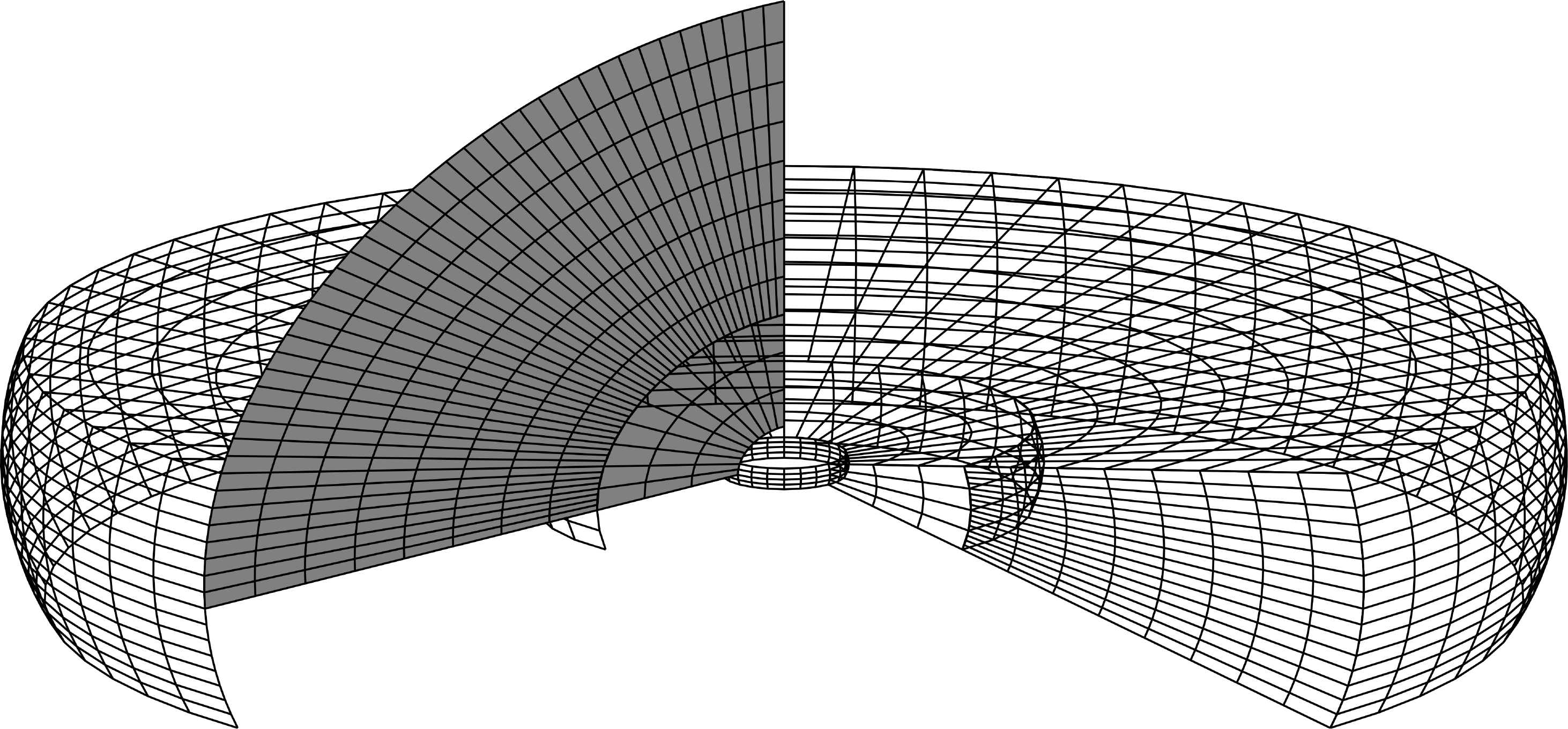}
    \caption{An illustrative picture of the spherical mesh of the equatorial region. Angular resolution doubles in both polar and azimuthal coordinates in the outer region. The figure is only for illustration, the resolution of our simulation is shown in Table \ref{tab:angular_res}.}
    \label{fig:mesh}
\end{figure}

$\tau(x,y,z)$ is calculated by interpolating $\tau(r,\theta,\phi)$ linearly after translating $(x,y,z)$ to the frame whose origin is the center of the primary star. $\tau(r,\theta,\phi)$ is the optical depth trace back to an angular dependent surface $r_{\rm{surf}}(\theta_{\rm{j}},\phi_{\rm{k}})$ that encloses the photosphere and probably the chromosphere of the primary star. For simplicity, $r_{\rm{surf}}$ is defined as the smallest contour of $\rho=10^{-10}$ \gcmc\ that encloses the AGB star. $\tau$ is defined on a discrete spherical coordinate
\begin{equation}
    \tau(r_{\rm{i}},\theta_{\rm{j}},\phi_{\rm{k}})=\left\{
    \begin{array}{ll}
    \tau_{\rm{inside}}     &     r_{\rm{i}}<r_{\rm{surf}}, \\
    \int_{r_{\rm{surf}}}^{r_{\rm{i}}}\kappa(r,\theta_{\rm{j}},\phi_{\rm{k}})\rho(r,\theta_{\rm{j}},\phi_{\rm{k}}) dr     &  r_{\rm{i}}\ge r_{\rm{surf}},
    \end{array}
    \right.
\end{equation}
where $\theta_\text{j}\in[0,\pi]$ is the discretized polar angle and $\phi_\text{k}\in[0,2\pi]$ is the discretized azimuthal angle. $\tau_{\rm{inside}}$ is a large number but is irrelevant to the actual calculation because $a_{\rm{rad}}$ is set to 0 within $r_{\rm{surf}}$. To distinguish, we will call a finite volume in Cartesian coordinate a 'cell' and a finite volume in spherical coordinate a 'bin' as shown in Figure \ref{fig:mesh}. The center of this spherical coordinate is at the center of the primary star therefore the coordinate is translated.
$\kappa(r_{\rm{i}},\theta_{\rm{j}},\phi_{\rm{k}})$ and $\rho(r_{\rm{i}},\theta_{\rm{j}},\phi_{\rm{k}})$ are mapping and regriding of the data in the translated Cartesian coordinate. To distinguish the translated Cartesian coordinate from the untranslated coordinate, we use $(x',y',z')$ to denote the translated Cartesian coordinate. The density of a bin in the spherical coordinate is defined as
\begin{equation}\label{eqn:rho_spherical}
    \rho(r_{\rm i},\theta_{\rm j},\phi_{\rm k})=\frac{\sum\rho(x',y',z')dv(x',y',z')\delta(x',y',z')}{\sum dv(x',y',z')\delta(x',y',z')},
\end{equation}
where $dv$ is the finite volume in the translated coordinate. To achieve a high precision of $\rho(r_{\rm i},\theta_{\rm j},\phi_{\rm k})$, we apply 4 levels of mesh refinement to all the quantities that are defined on the cells, that means each $dv$ has a physical size of $(0.05\rm{AU})^3$. $\delta(x',y',z')$ can be thought of as a kernel function, i.e.,
\begin{equation}\label{eqn:kernel}
    \delta(x',y',z')\left\{
	\begin{array}{ll}
		1  & \mbox{if } (x',y',z')\in dv(r_\text{i},\theta_\text{j},\phi_\text{k}), \\
		0 & \mbox{if } (x',y',z')\not\in dv(r_\text{i},\theta_\text{j},\phi_\text{k}).
	\end{array}
\right.
\end{equation}
The membership relation $\in$ can be defined explicitly as,
\begin{eqnarray}
    r_\text{i-1/2}&\le& r<r_\text{i+1/2},  \\
    \theta_\text{j-1/2}&\le&\theta<\theta_\text{j+1/2},  \\
    \phi_\text{k-1/2}&\le&\phi<\phi_\text{k+1/2},  \label{eqn:phi}
\end{eqnarray}
where $(r,\theta,\phi)$ is the spherical coordinate of $(x',y',z')$.

Similarly, we define the opacity of a bin as mass weighted average
\begin{align}
    &\kappa(r_\text{i},\theta_\text{j},\phi_\text{k})=\nonumber  \\
    &\frac{\sum\rho(x',y',z')\kappa(x',y',z')dv(x',y',z')\delta(x',y',z')}{\sum\rho(x',y',z')dv(x',y',z')\delta(x',y',z')}.
\end{align}
In the mapping and regriding process, the computational domain is divided into two regions, one is the polar region and the other is the equatorial region. The polar region does not resolve the azimuthal angle, therefore, in the polar region, Equation \ref{eqn:kernel} becomes
\begin{equation}\label{eqn:kernelpolar}
    \delta(x',y',z')\left\{
	\begin{array}{ll}
		1  & \mbox{if } (x',y',z')\in dv(r_\text{i},\theta_\text{j}), \\
		0 & \mbox{if } (x',y',z')\not\in dv(r_\text{i},\theta_\text{j}).
	\end{array}
\right.
\end{equation}
The equatorial region resolves the azimuthal angle. Angular resolution is increased for both two regions as $r$ gets larger. We illustrate the spherical grid that is used for the radiative transfer in Figure \ref{fig:mesh}. The resolution structure that we use is listed in Table \ref{tab:angular_res}.

\begin{table}[ht!]
    \centering
    \begin{tabular}{|c|c|c|c|c|c|}
    \hline
    \multicolumn{3}{|c|}{Inner region}
    &  \multicolumn{3}{c|}{Refined region} \\ \hline
    \multicolumn{3}{|c|}{$r\in[1.25,2.5]$ AU}
    &  \multicolumn{3}{c|}{$r\in[2.5,36]$ AU} \\ \hline
    $\rm{d}r$  &  $\rm{d}\theta$ &  $\rm{d}\phi$  &  $\rm{d}r$  &  $\rm{d}\theta$  &  $\rm{d}\phi$ \\ \hline
    $0.05$AU  &  $\pi/50$  &  $\pi/50$  &  $0.05$AU  &  $\pi/100$  &  $\pi/100$ \\ \hline
    \end{tabular}
    \caption{Angular resolution of the equatorial region in radiative transfer. The polar region has the same angular resolution except that azimuthal angle is not resolved.}
    \label{tab:angular_res}
\end{table}

In Section \ref{sec:dustformation}, we calculate the dust-free optical depth by setting $\kappa=\kappa_{\rm{mol}}=2.5\times10^{-4}$ \gcmc\ everywhere. The $\tau(r,\theta,\phi)$ for the radiation-hydrodynamic simulation is re-calculated by using the adopted opacity profile in Section \ref{sec:opacity}.

\end{document}